\documentclass{mn2e}
\usepackage{times}
\usepackage{amsfonts}
\usepackage{graphicx}

\title[]{Log-Poisson Non-Gaussianity of Ly$\alpha$ Transmitted Flux
   Fluctuations at High Redshift}

\author[Lu, Zhu, Chu, Feng \& Fang]
{Yi Lu$^{1}$\thanks{E-mail:lulu@mail.ustc.edu.cn},
Weishan Zhu$^{2,3}$, Yaoquan Chu$^{1}$, Long-long Feng$^{3}$,
Li-Zhi Fang$^{2}$\\
$^{1}$Center for Astrophysics, University of Science
and Technology of China, Hefei, Anhui 230026, China\\
$^{2}$Department of Physics, University of Arizona,
Tucson, AZ 85721\\
$^{3}$Purple Mountain Observatory, Nanjing, 210008, China}
\begin{document}

\pagerange{\pageref{firstpage}--\pageref{lastpage}} \pubyear{2010}

\maketitle

\label{firstpage}

\begin{abstract}

We investigate the non-Gaussian features of the IGM at redshift $z\sim 5
- 6$ using Ly$\alpha$ transmitted flux of quasar absorption spectra
and cosmological hydrodynamic simulation of the concordance $\Lambda$CDM
universe. We show that the neutral hydrogen mass density field and Ly$\alpha$
transmitted flux fluctuations possess all the non-Gaussian features predicted by
the log-Poisson hierarchy, which depends only on two dimensionless
parameters $\beta$ and $\gamma$, describing, respectively, the
intermittence and singularity of the random fields. We find that the
non-Gaussianity of the Ly$\alpha$ transmitted flux of quasars from
$z=4.9$ to $z=6.3$ can be well reconstructed by the hydrodynamical
simulation samples. Although the Gunn-Peterson optical depth and its variance
underwent a significant evolution in the redshift range of $5 - 6$, the
intermittency measured by $\beta$ is almost redshift-independent in this range.
More interesting, the intermittency of quasar's absorption spectra on physical scales
$0.1-1$ h$^{-1}$Mpc in redshift $5 - 6$ are found to be about the same
as that on physical scales $1-10$ h$^{-1}$Mpc at redshifts $2 - 4$.
Considering the Jeans length is less than 0.1 h$^{-1}$Mpc at $z\sim 5$,
and $1$ h$^{-1}$Mpc at $z\sim 2$, these results imply that the nonlinear
evolution in high and low redshifts will lead the cosmic baryon fluid to a state
similar to fully developed turbulence. The log-Poisson high order behavior of current 
high redshift data of quasar's spectrum can be explained by uniform UV background 
in the redshift range considered. We also studied the log-Poisson non-Gaussianity 
by considering inhomogeneous background. With several simplified models of 
inhomogeneous background, we found the effect of the inhomogeneous background on 
the log-Poisson non-Gaussianity is not larger than 1-sigma. 

\end{abstract}


\section[]{Introduction}

The cosmic density and velocity fields of the intergalactic medium(IGM) at low redshifts are
highly non-Gaussian, while the temperature fluctuations of the
cosmic microwave background radiation basically are Gaussian or at
most weakly non-Gaussian. Thus, the non-Gaussian features of the IGM at
low redshifts should mainly result from nonlinear dynamical processes.
A well knowledge of these non-Gaussian features and their development history
can fundamentally improve our understanding of the formation
and evolution of the structures in the universe. For instance, the lognormal
clustering model of the IGM (Bi, 1993; Bi \& Devidson
1997; Liu et al 2006; Feng et al 2008) can explain well all the basic
properties of the Ly$\alpha$ forests of quasar's absorption spectrum at various
redshifts. However, high order statistics do reveal the deviation of
observation from the lognormal model (Jamkhedkar et al. 2003; Lu et al 2009).

The aim of this paper is to study the non-Gaussian features of Ly$\alpha$ transmission
flux of quasar spectrum at high redshift. The non-Gaussianities of high redshift
Ly$\alpha$ transmission flux have been studied by various statistics, including
the probability distribution function (PDF) of the flux (Fan et al. 2002;
Becker et al. 2007), the distribution of the size of dark gaps (Songaila \&
Cowie 2002; Fan et al 2006), the largest peak width distribution (Gallerani et
al 2007) and the distributions of the width of leaks (Liu et al 2007; Feng et al
2008). Most of these statistics are designed based on the following observation
facts: the high redshift quasar absorption spectra consist of
complete absorption troughs separated by the spikes of transmitted
flux. Obviously, these statistics are not suitable for comparing the
non-Gaussian features at high redshift and low redshift, as
the absorption spectra of low redshift quasars consist of Ly$\alpha$
forests and do not contain Gunn-Peterson troughs. Our focus is on
the log-Poisson non-Gaussianity, which are characterized by two
dimensionless parameters that are available for both high and low redshift
samples. Therefore, it can be a useful tool to study the redshift evolution of
non-Gaussianity.

More important, log-Poisson hierarchy is directly related to the
dynamics of cosmic baryon fluid. With the cosmological hydrodynamical
simulation of the concordance $\Lambda$CDM model, it has been shown
that, in the scale range from the onset of nonlinear evolution to the scale
of dissipation, the velocity fields of cosmic baryon fluid at low redshift
are extremely well described by the She-Leveque's scaling formula (He et al 2006),
which was inferred from the log-Poisson hierarchical cascade (Dubrulle 1994; She \&
Wamire 1995; Benzi et al. 1996). The non-Gaussian behavior of the mass
density field of the baryon fluid also can be well described by log-Poisson
processes (Liu \& Fang 2008). Recently, high order statistics of observed high
resolution and high S/N Ly$\alpha$ absorption spectra of quasars at redshift
$z\sim$ 2 - 3 are found to be well consistent with the non-Gaussian features predicted
by the log-Poisson hierarchy (Lu et al 2009). It would be worth
to study whether the scenario of self-similar log-Poisson hierarchical cascade
is still hold at high redshifts. Whether the log-Poisson non-Gaussianity experience
a strong evolution in the redshift range 5 - 6? Can we explain the log-Poisson
non-Gaussianity and its redshift dependence with the concordance $\Lambda$CDM
model?

A possible application of the non-Gaussianity of neutral hydrogen
distribution is to constrain the fluctuations of hydrogen-ionizing
radiation background field. Considering the inhomogeneity of UV
background might be a source of the non-Gaussianity of ionized
 and neutral hydrogen, the non-Gaussianity of
Ly$\alpha$ transmitted flux would shed light on the evolution of UV
background. The inhomogeneity of UV background at lower redshifts
are negligible and a uniform ionizing background is an reasonable
approximation. The situation becomes more complex and debatable when
$z > 5$ (Fan et al. 2002; Wyithe \& Loeb 2005; Liu, et al. 2006; Fan
at al. 2006; Liu et al. 2007; Mesinger \& Furlanetto, 2009; Furlanetto 
\& Mesinger 2009).
Therefore, it is also worth to study the effect of non-uniform UV 
background on the log-Poisson high order behavior at high redshifts. 

This paper is organized as follows. \S 2 gives the theoretical
background. \S 3 presents the log-Poisson hierarchy of Ly$\alpha$
transmitted flux fluctuations of observed samples of high redshift
quasar absorption spectra. In \S 4, we describe the method used for
producing simulation samples, and show the log-Poisson
non-Gaussianity of the neutral hydrogen component of cosmic baryon
fluid. \S 5 compares the log-Poisson non-Gaussianities of observed
data with simulation samples of Ly$\alpha$ transmitted flux. The
effect of non-uniform UV background on non-Gaussianity is also
studied in \S 5. Finally, conclusion and discussion are given in \S
6.

\section[]{Theoretical background}

\subsection{Hierarchical clustering}

It has been recognized for a long time that the clustering process
of cosmic matter probably is hierarchical (Peebles 1980). That is,
the nonlinear dynamics of large scale structure formation can be
characterized by the merging of holes from smaller to larger
scales. An early model of hierarchical clustering assumes that the
correlation functions of the density field satisfy a linked-pair
relation, $\zeta_n=Q_n\zeta_2^{n-1}$ (White, 1979), where $\zeta_n$
is the $n$th irreducible correlation function with variable $\delta
\rho({\bf x})=\rho({\bf x})-\bar{\rho}$, where $\rho({\bf x})$ and
$\bar{\rho}$ are, respectively, the cosmic mass density field and
its mean.  However, observation samples of transmitted flux of
Ly$\alpha$ forest do not support the linked-pair relation if the
coefficients $Q_n$ are assumed to be constants (e.g. Feng et al.
2001).

The hierarchical merging has also been modeled by an additive
cascade rule (Cole and Kaiser, 1988). The basic step of the cascade
rule is to assume that the a cell of mass $M$ and  spatial scale $x$
will evolve into two cells, 1 and 2 with mass $M_1=M+\delta m$ and
$M_2=M-\delta m$, on scale $x/2$, where $\delta m$ is a Gaussian
random variable. However, central limit theorem shows that the field
produced by an additive cascade process should be Gaussian. It
cannot explain the non-Gaussian features of Ly$\alpha$ transmitted
flux (Pando et al. 1998). Moreover, this cascade needs a ''very
small initial units'' (Peacock, 1999) as the first generation halo
of the hierarchy, of which the mass is non-zero. This ''initial
units'' is not compatible with hydrodynamic equations with
continuous variables.

A proper model of the the hierarchical clustering should be randomly
multiplicative, and infinitely divisible. For a randomly
multiplicative cascade, the mass $m_n$ in a cell at step $n$ is
related to step $n-1$ by $m_n=(1\pm \delta)m_{n-1}$, where $\delta$
is a random variable. Non-Gaussian features can be formed through
randomly multiplicative cascade processes, even if the original
field is Gaussian. The infinite divisibility means that there is no
finite ''initial units'' in the hierarchical merging. The ''initial
units'' of the merging process can be infinitesimal.

The log-Poisson hierarchy has all these desired properties. More
important, the log-Poisson hierarchy actually was inferred from the
invariance and symmetry of the Navier-Stokes equations and works
well in explaining the high order behavior of fully developed
turbulence (Dubrulle, 1994; She \& Waymire 1995; Leveque \& She
1997).

\subsection{Log-Poisson hierarchy}

To measure the non-Gaussianity caused by log-Poisson hierarchy, it
is better to use the variable $\delta\rho_{r}= \rho({\bf
x+r})-\rho({\bf x})$, $r=|{\bf r}|$, but not $\rho({\bf
x+r})-\bar{\rho}$. For a statistically isotropic and homogeneous
random field, one can just consider  $|\delta\rho_{r}|$, as the
distribution of positive and negative $\delta\rho_{r}$ is
statistically symmetric. The basic statistical quantity is the
structure function defined by
\begin{equation}
S_p(r)\equiv \langle |\delta\rho_r|^p\rangle,
\end{equation}
where  $p$ is the order of statistics, and the average $\langle ...\rangle$
is taken over the ensemble of density fields. The second-order structure function
$S_2=\langle |\delta\rho_r|^2\rangle$ as a function of $r$ (scale) is actually
the power spectrum of the mass density field (Fang \& Feng 2000).

In scale-free range, the structure function should be a function of
power law of $r$ as
\begin{equation}
S_p(r)\propto r^{\xi(p)},
\end{equation}
where $\xi(p)$ is referred to intermittent exponent. For a Gaussian
field, $\xi(p)$ is a linear function of $p$, but for a intermittent
field, function $\xi(p)$ is nonlinear.

The log-Poisson hierarchy assumes that, in the scale-free range, the
variables $\delta\rho_{r}$ on different scales $r$ are related to
each other by a statistically hierarchical relation as (Dubrulle
1994; She \& Waymire 1995)
\begin{equation}
\delta\rho_{r_2} = W_{r_1r_2}\delta\rho_{r_1}.
\end{equation}
The factor $W_{r_1r_2}$ is a function of the ratio $r_1/r_2$ given by
\begin{equation}
W_{r_1r_2}=\beta^m (r_1/r_2)^{\gamma},
\end{equation}
which describes how the fluctuation $\delta\rho_{r_1}$ on the
larger scale $r_1$ relates to fluctuations $\delta\rho_{r_2}$ on
the smaller scale $r_2$. In eq.(4), $m$ is a Poisson random variable
with the PDF
\begin{equation}
P(m)=\exp(-\lambda_{r_1r_2})\lambda_{r_1r_2}^m/m!.
\end{equation}
The random variable $m$ can be considered as the steps of the
evolution from $\delta\rho_{r_1}$ to $\delta\rho_{r_2}$. To ensure
the normalization $\langle W_{r_1r_2} \rangle=1$, where the average
$\langle...\rangle$ is over $m$, the mean $\lambda_{r_1r_2}$ of the
Poisson distribution is then
\begin{equation}
\lambda_{r_1r_2}= \gamma[\ln(r_1/r_2)]/(1-\beta).
\end{equation}
The log-Poisson hierarchy contains both the spatial size and
amplitude of the density fluctuations. This point is different from
other hierarchical models, which consider only the size of
hierarchical units.

The log-Poisson hierarchy given by eq.(3) depends only on the ratio
$r_1/r_2$, which is obviously scale invariant. The hierarchy is
determined by two dimensionless positive parameters: $\beta$ and
$\gamma$, describing, respectively, the intermittence and
singularity of the random fields. Equation (3) relates
$\delta\rho_{r}$ on different scales by multiplying a random factor
$W$, which generally yields a non-Gaussian field even if the field
originally is Gaussian (Pando et al. 1998).

The cascade from scale $r_1$ to $r_2$, and then to $r_3$ is
identical to the cascade from $r_1$ to $r_3$. It is because
$W_{r_1r_3}=W_{r_1r_2}W_{r_2r_3}=\beta^N (r_1/r_3)^{\gamma}$, where
$N$ is again a Poisson random variable with
$\lambda_{r_1r_3}=\lambda_{r_1r_2}+\lambda_{r_2r_3}=\gamma[\ln(r_1/r_3)]/(1-\beta)$.
The log-Poisson hierarchy removes an arbitrariness in defining the
steps of cascade from $r_1$ to $r_2$ or $r_2$ to $r_3$. Therefore,
the log-Poisson hierarchy, suggested by eq.(3), is discrete in terms
of the discrete random number $m$. However, the scale $r$ is
infinitely divisible. Namely, there is no lower limit on the
difference $r_1-r_2$. It can be infinitesimal, and the hierarchical
process is of infinite divisibility. With the log-Poisson model
eqs.(3)-(6), one can show that the intermittent exponent $\xi(p)$ is
given by (Liu \& Fang 2008)
\begin{equation}
\xi(p)=-\gamma[p-(1-\beta^{p})/(1-\beta)].
\end{equation}
This is actually the SL scaling formula (She \& Leveque 1994).

From eq.(7), the power spectrum $S_2(r)={\rm const}$ is flat, or
called white. However, the power spectrum of the initial Gaussian
field of the cosmic matter generally is not white, but colored with
power-law $S_2(r)\propto r^{-2\alpha}$. In this case, we should
adjust the log-Poisson hierarchy eq.(3) by replacing
$\delta\rho_{r_1}$ and $\delta\rho_{r_2}$ with
$r_1^{\alpha}\delta\rho_{r_1}$ and $r_2^{\alpha}\delta\rho_{r_2}$
respectively. The intermittent exponent $\xi(p)$ is
\begin{equation}
\xi(p)=-\alpha p-\gamma[p-(1-\beta^{p})/(1-\beta)].
\end{equation}
When parameter $\alpha=0$, eq.(8) will simplify to eq.(7).
The non-Gaussian features of the field described by the log-Poisson
hierarchical clustering have been given in Liu \& Fang (2008) and Lu
et al (2009).

\section[]{Log-Poisson non-Gaussianity of observed samples}

\subsection{Observed data}

Observational data used here consists of the spectra of 19 QSOs with
redshifts from $z=$5.74 to 6.42 that compiled in Fan et al. (2006).
The data have resolution of $R \sim 3000-4000$ and are re-binned to
a resolution $R=2600$. The observed flux, $f_{\rm obs}$, is
normalized with a power-law continuum $f_{\rm con}\propto
\nu^{-0.5}$. The noise and continuum uncertainty of transmitted flux
$\mathbb{F} \equiv f_{\rm obs}/f_{\rm con}$ is in the level of
$0.018 \pm 0.012$ in the range $z\leq 5.7$, and $0.014\pm 0.008$ in
the range $z>5.7$ (Liu et al. 2007). For more details, we refer to
Fan et al. (2006). Wavelength of the data covers roughly the
rest-frame wavelength from 900 to 1350\AA. To avoid the mixing of
Ly$\beta$ absorption and the proximity effect, only pixels have the
rest frame wavelength $>1040$ \AA \ and below the maximum Ly$\alpha$
are used.

In our analysis below, we use only the  Ly$\alpha$ transmitted flux
in the redshift range from 4.7 to 6.3. There actually are 12 quasars
available in the range  $z > 5.9$ and reduce further to 4 in $6.1 < z< 6.3$. We
divide the redshift range from 4.9 to 6.3 into 7 bins by $4.9 +
n\times 0.20 < z < 4.9 + (n+1)\times 0.20$ where $n=0, 1, \ldots 6$.
More specifically, the redshift size of  each bin is $\Delta z=0.20$. As a
comparison, we also use sample in the redshift range from 4.7 to
6.2, and divide it into 10 bins by $4.7 + n\times 0.15 < z < 4.7
+ (n+1)\times 0.15$ where $n=0, 1, \ldots 9$; each bin
has a size of $\Delta z= 0.15$. All the transmission flux pixels in a given
redshift bin form an ensemble. The numbers of pixels in
different ensemble are not uniform.

In order to compare with observation at moderate redshift we also analyze
the high resolution and high signal to noise ratio Ly$\alpha$ absorption spectra
 of 28 Keck High Resolution (HIRES) QSO (Kirkman \& Tytler 1997). It is
 the same as that used in Lu et al (2009). The details of the data set
 and its reduction have been described  in Jamkhedkar et al. (2002, 2003).

With these samples we calculate the optical depth $\tau(z)=-\ln\mathbb{F}(z)$,
and the fluctuation of the optical depth $\delta\tau_r=\tau(x+r)-\tau(x)$,
where the spatial coordinates $x$ and $r$ are in physical scale. Since the
variable $\delta\tau_r$ is given by difference between $\tau(x+r)$ and $\tau(x)$,
$\delta\tau_r$ is independent of fluctuations of $\tau(x)$ on scales larger than $r$.
Therefore, the variable $\delta\tau_r$ actually is insensitive to the continuum
used in the data reduction (Jamkhedkar et al. 2001).

\begin{figure*}
\center
\vspace{1cm}
\includegraphics[scale=0.40,angle=-90]{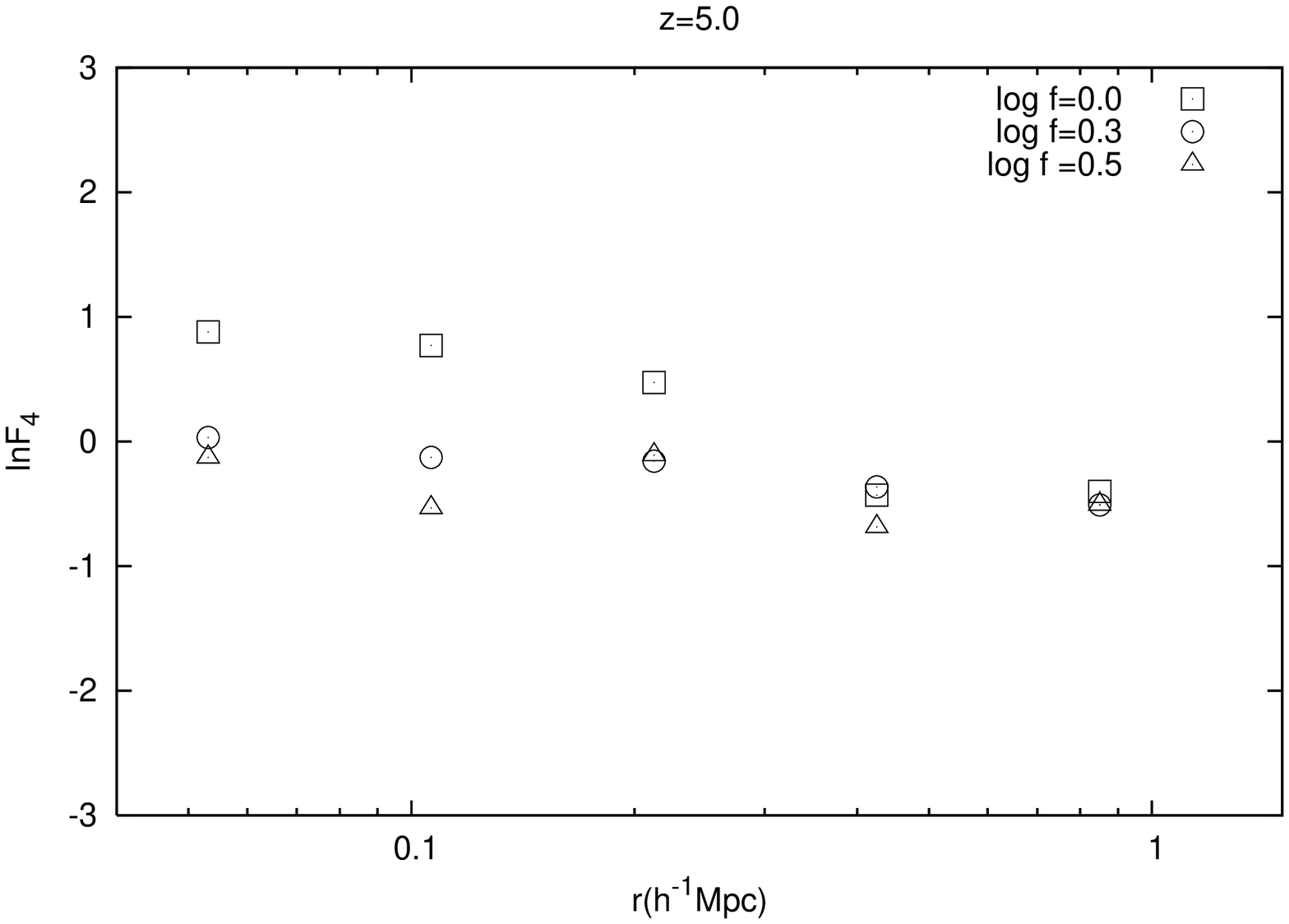}
\includegraphics[scale=0.40,angle=-90]{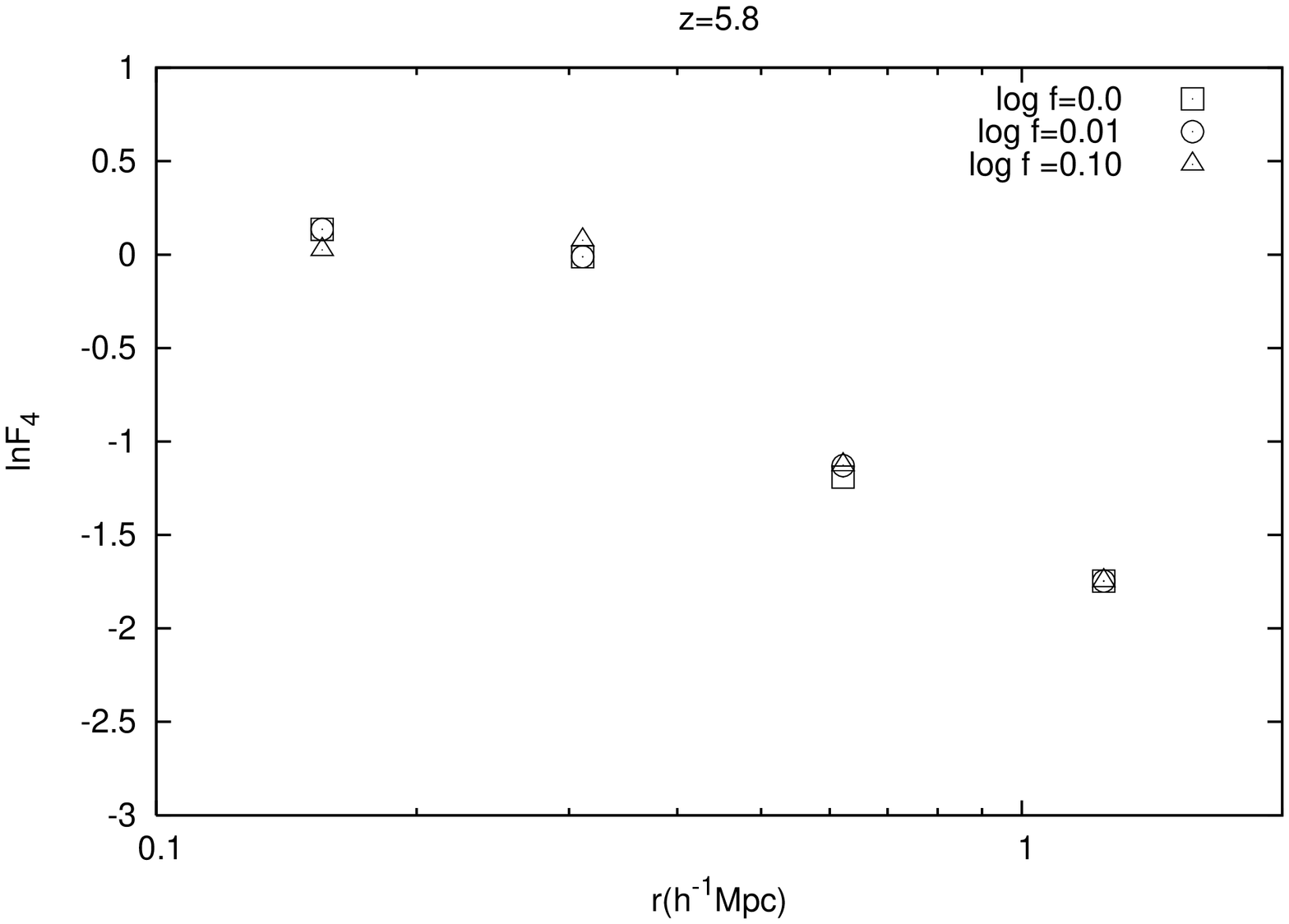}
\caption{$F_4(r)$ given by observed sample of the Ly$\alpha$
transmitted flux at redshift ranges $4.9< z < 5.1$ (left), and
$5.7<  z < 5.9$ (right).}
\end{figure*}

As has been shown in Lu et al (2009), the variable $\delta\tau_r$ is approximately
a measurement of the density fluctuation $\rho_r$. We can then study
the log-Poisson non-Gaussianity with variable $\delta\tau_r$. For
instance, the structure function with the variable $\delta\tau_r$ is
defined now by
\begin{equation}
S_p(r)\equiv \langle |\delta\tau_r|^p\rangle.
\end{equation}

To treat the unwanted data, including low S/N and bad pixels, we use the algorithm of
wavelet denoising by threshold (Donoho 1995, Jamkhedkar, et. al. 2003).
This method is effective for pixelated data. First, we calculate the
wavelet scaling function coefficients (SFCs) of both transmission
flux field $\mathbb{F}(x)$ and noise field $n(x)$ with
\begin{equation}
\epsilon_{jl}^{\mathbb{F}}=\int\mathbb{ F}(x)\phi_{j,l}(x)dx ,
\hspace{3mm} \epsilon_{jl}^N=\int
n(x)\phi_{j,l}(x)dx.
\end{equation}
where $\phi_{j,l}(x)$ is the scaling function of wavelet on scale
$j$ and at position $l$. We then identify unwanted mode $(j,l)$ by using
the threshold condition
\begin{equation}
|\epsilon_{jl}^{\mathbb{F}}/\epsilon_{jl}^N| < f.
\end{equation}
This condition flags all modes with S/N less than $f$. We skip all
the flagged modes when doing statistics. To reduce the boundary
effect of unwanted chunks, we also flag two models around an
unwanted model. With this method, no rejoining and smoothing of the
data are needed. The threshold $f$ is given by the same way as
Jamkhedkar et al (2003) and Lu et al (2009).

A typical statistical quantity of log-Poisson non-Gaussianity is
\begin{equation}
F_p(r)\equiv S_{p+1}(r)/S_p(r).
\end{equation}
We test the effect of noise on
$F_4(r)\equiv S_5(r)/S_4(r)$.  We calculate $F_4(r)$ for data sets
given by different threshold $f$. The $f$-dependence of $F_4(r)$ is
shown in Figure 1. For the data set at $4.9< z <5.1$, the values of
$F_4(r)$ are $f$-independent when $\log f\geq 0.3$. In other words,
the statistical results are stable with respect to threshold $f
\simeq 2$. We will use only data with S/N larger than 2. This
threshold is larger than the error level given by Fan et al (2006).
For sample set of $5.7 <z <5.9$ (right panel of Figure 1), $F_4(x)$
are very weakly dependent on $\log f$. It is because, in the
redshift range $z >5.7 $, the high order statistics $F_4(x)$ are
dominated by modes with high $\delta\tau_r$, and therefore, it
is insensitive to dropping modes with low $\delta\tau_r$. However, the
fewer the modes, the larger the variance of Poisson process.
Therefore, we should consider the variance of Poisson process in
our statistics.

\subsection{Redshift dependence of parameter $\beta$}

\begin{figure*}
\center
\includegraphics[scale=0.35,angle=-90]{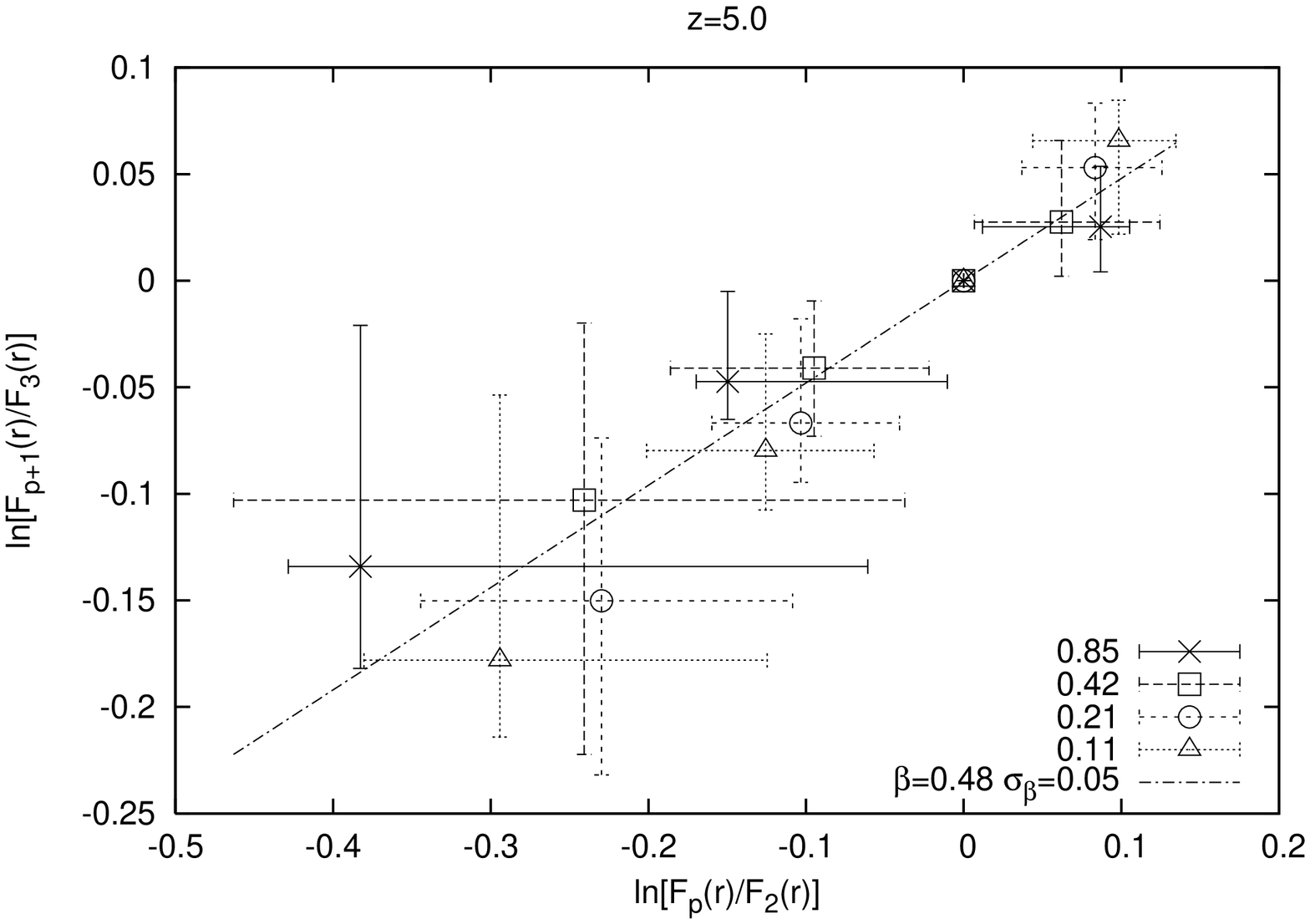}
\includegraphics[scale=0.35,angle=-90]{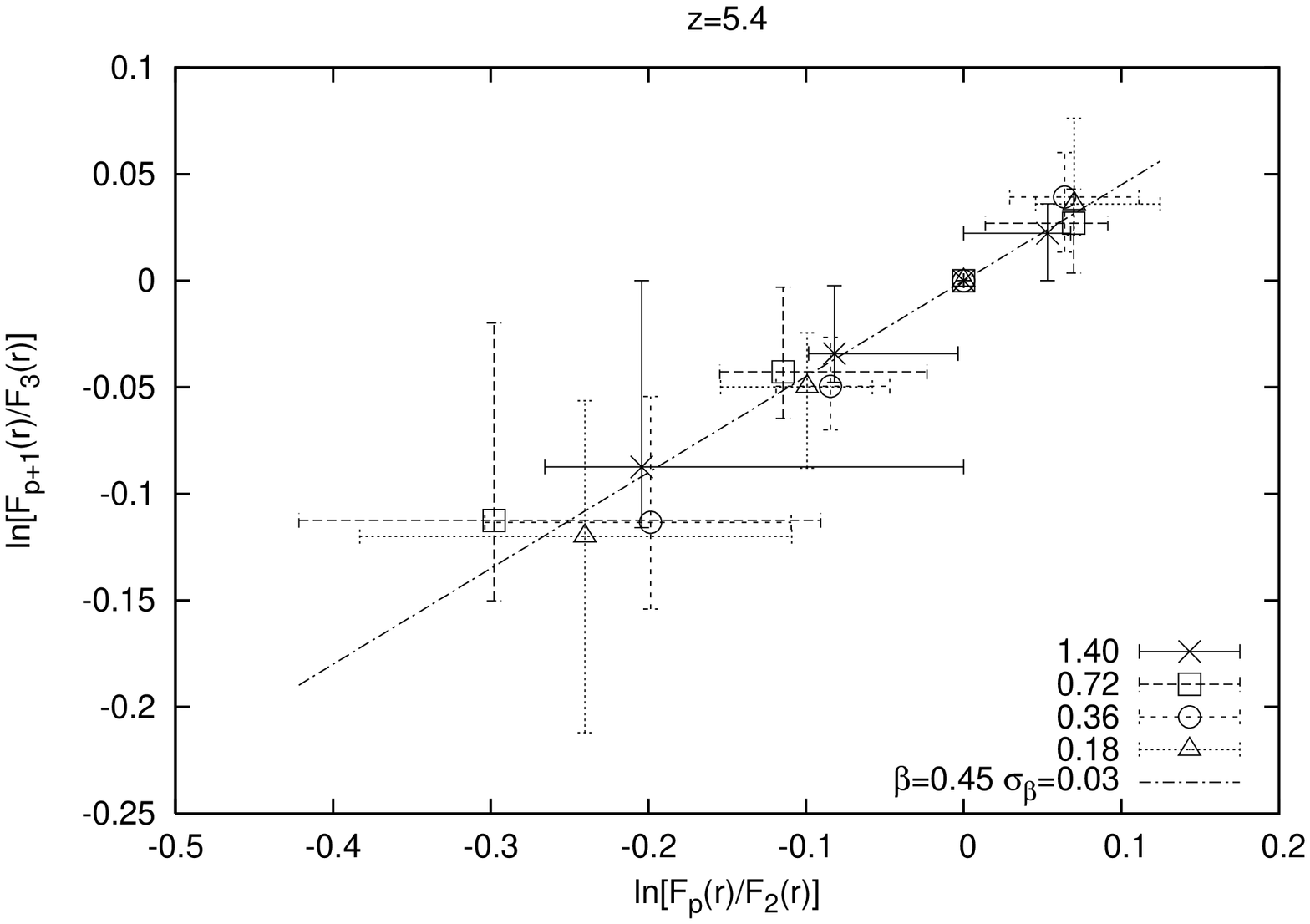}
\includegraphics[scale=0.35,angle=-90]{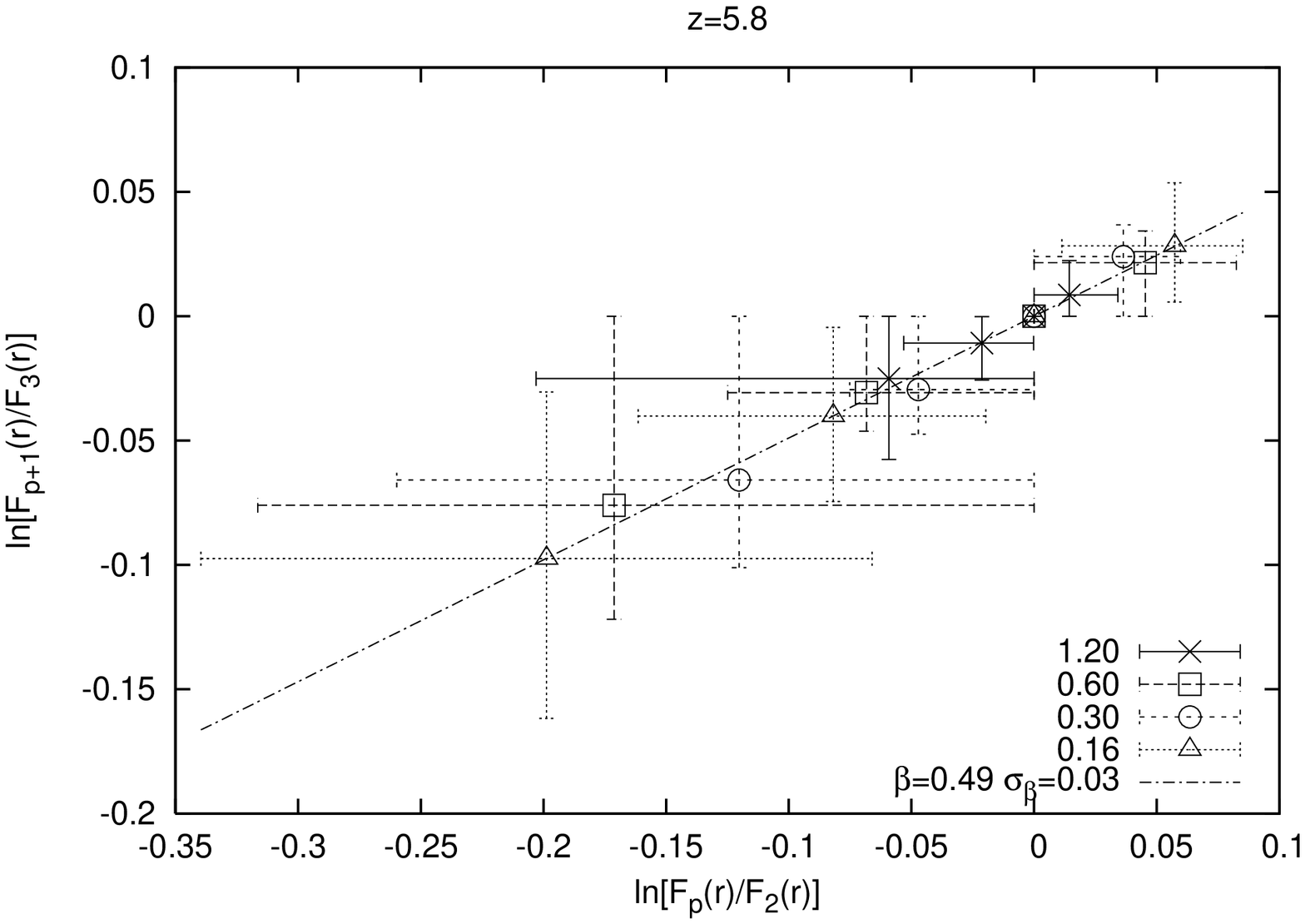}
\includegraphics[scale=0.35,angle=-90]{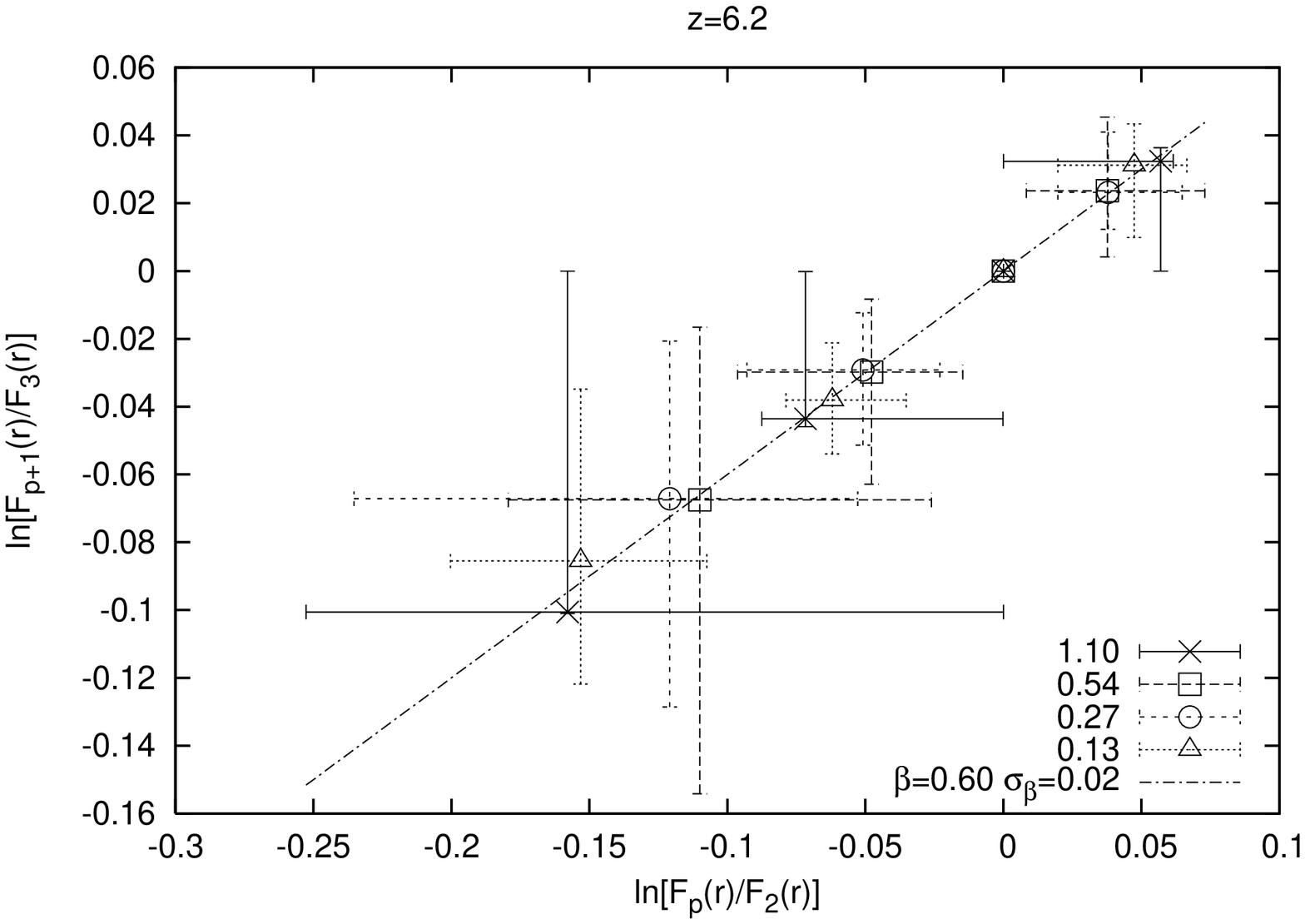}
\caption{The $\beta$-hierarchy of observed sample of the Ly$\alpha$
transmitted flux at redshift $z=5.0$ (top-left) , 5.4 (top-right),
5.8 (bottom-left) and 6.2 (bottom-right). The physical scale $r$ is
in the range $\sim$ 0.1 - 1.5 h$^{-1}$ Mpc, and order $p$ is from 1
to 2.5 . The error bars are given by the maximum and minimum of
bootstrap re-sampling.}
\end{figure*}

We first study the $\beta$-hierarchy predicted by the log-Poisson hierarchy.
It reads (Liu \& Fang, 2008)
\begin{equation}
\ln F_{p+1}(r)/F_3(r)=\beta\ln F_{p}(r)/ F_2(r).
\end{equation}
Eq.(13) requires that for {\it all} $r$ and $p$, $\ln
[F_{p+1}(r)/F_3(r)]$ vs. $\ln [F_{p}(r)/F_2(r)]$ should be on a
straight line with slope $\beta$. which is called $\beta$-hierarchy.
Eq.(13) does not contain parameters $\gamma$ and $\alpha$. Figure 2
presents the $\beta$-hierarchy of observed transmitted flux in 4
redshift ranges $z=5.0$, 5.4, 5.8 and 6.2. The statistical quantity
$F_{p+1}(r)$ are given by all available observed data, of which the
physical scale $r$ covers the range from $\sim$ 0.1 to 1.5 h$^{-1}$
Mpc, and the order parameter $p$ increases from 1 to 2.5. All the
distributions of $F_{p+1}(r)/F_3(r)$ vs. $F_{p}(r)/ F_2(r)$ in
Figure 2 can be well fitted with a straight line. It shows that the
Ly$\alpha$ transmitted flux of real observation at high redshifts
still satisfy the $\beta$-hierarchy and indicates that the
log-Poisson non-Gaussian features are significant in the considered
redshift range.

\begin{figure*}
\center
\includegraphics[scale=0.38,angle=-90]{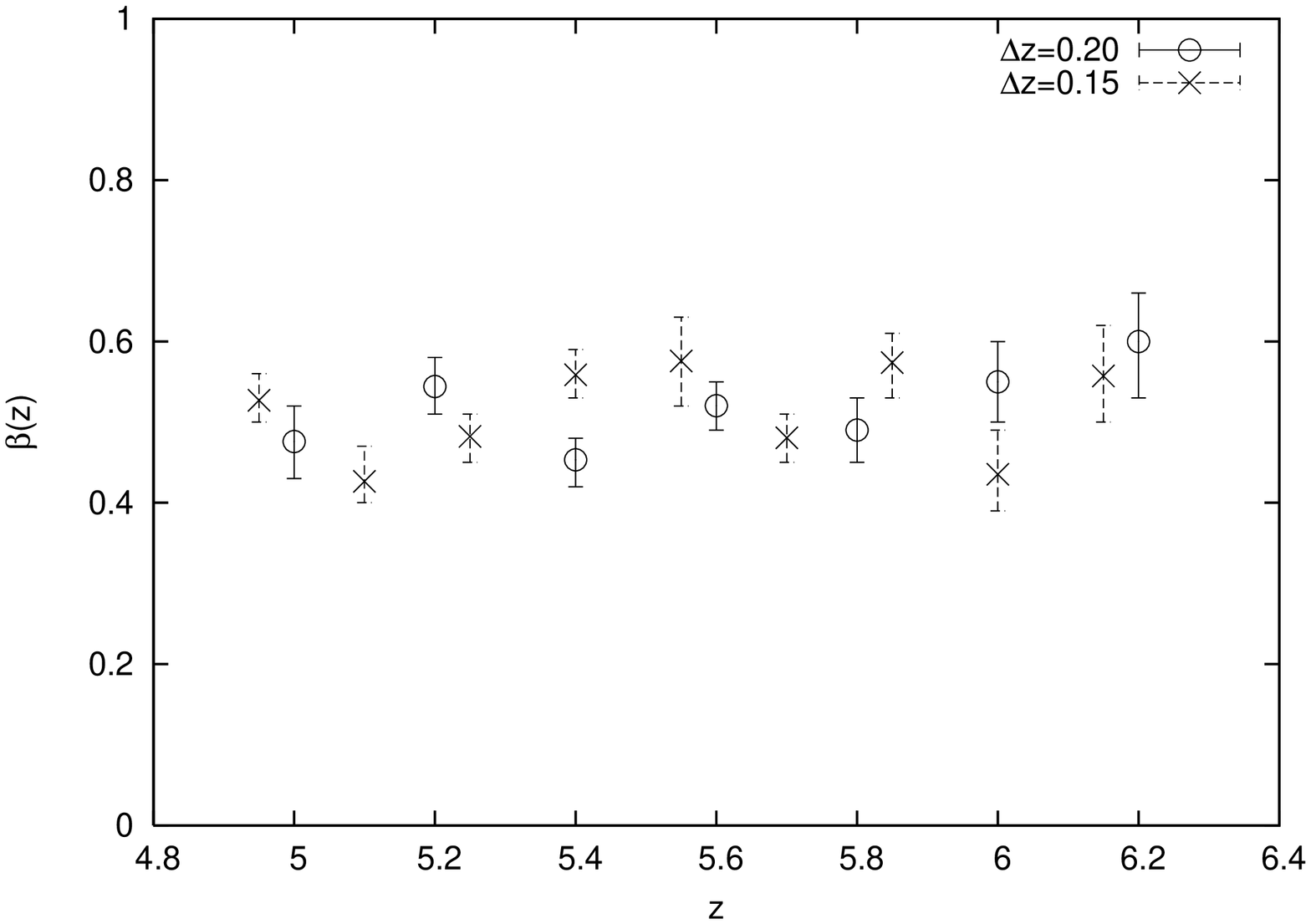}
\includegraphics[scale=0.38,angle=-90]{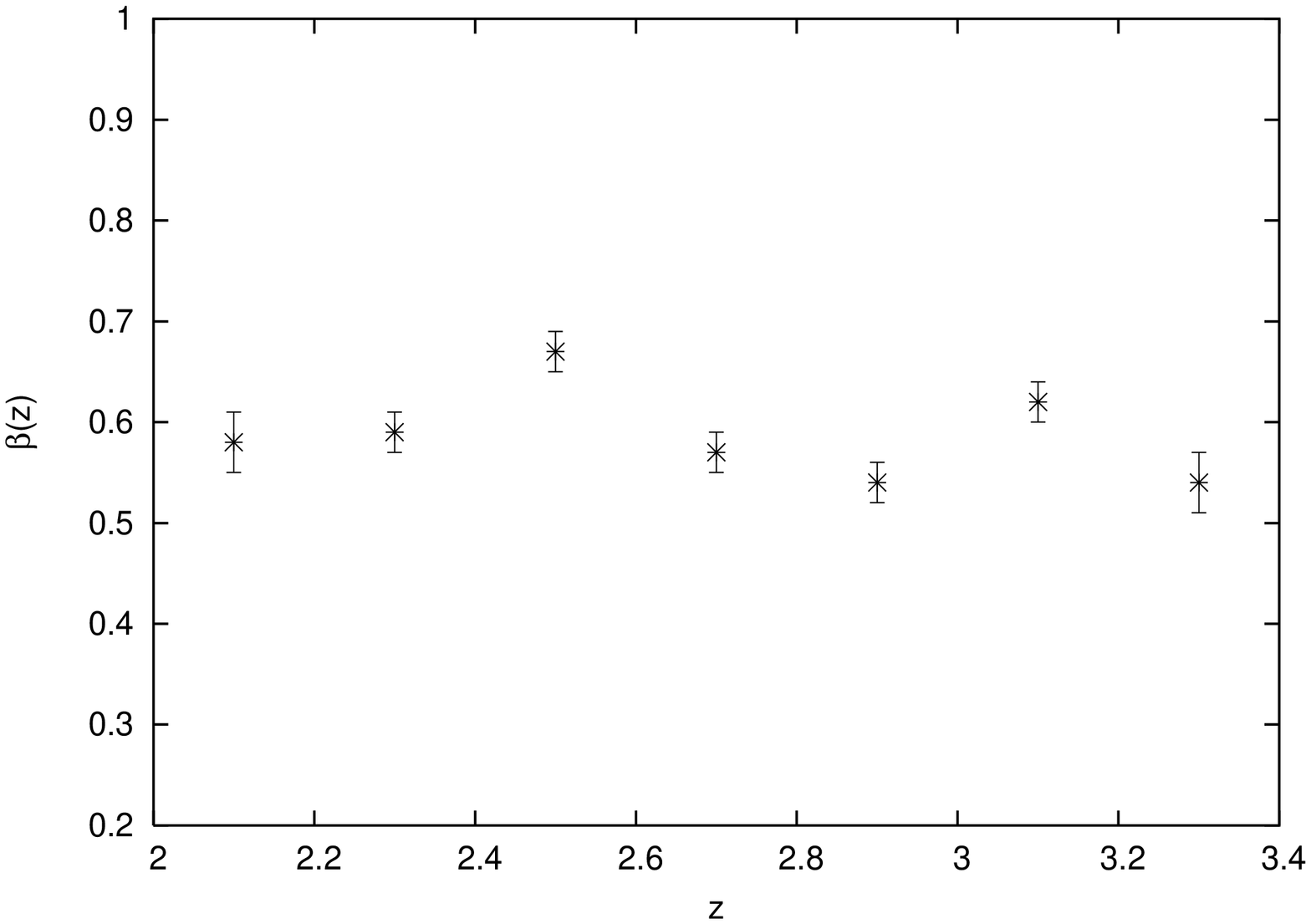}
\caption{Redshift dependence of parameter $\beta$ for samples of a.) Fan et al
(2006) (left), and b.) the Keck HIREs quasars, Jamkhedkar et al.(2003) (right).}
\end{figure*}

From the $\beta$-hierarchy straight line, we calculate the parameter
$\beta$ in redshift ranges from $z=5.0$ to 6.2.  Figure 3(a)
presents the mean of $\beta$ in each redshift bin. The error bars
are given by the variance of Poisson process, which generally are
large than the variance of $\beta$ in the given redshift bin. Figure
3(a) shows that the redshift evolution of $\beta$ is rather weak in
the range from $z=5$ to 6. We also calculate the redshift-dependence
of $\beta$ with the same data, but the size of redshift bin is taken
to be $\Delta z =0.15$. The result is also plotted in Figure 3(a),
which shows the same redshift dependence as that of $\Delta z
=0.20$. Therefore, the weak redshift-evolution of parameter $\beta$
does not affected by $\Delta z$. From now on, we will only give
results with $\Delta z=0.20$.

Figure 3(b) plots the redshift dependence of $\beta$ for the Keck
HIRES quasars spectra sample. The error bars are given by the
maximum and minimum in each redshift interval. Comparing Figure 3(a)
and (b), we can conclude that the non-Gaussian parameter $\beta$
evolves weakly from redshift 2 to 6. It should be pointed out that
the physical scale $r$ in Figure 3(a) (high redshift sample) is
actually smaller than that of Figure 3(b) (low redshift sample).
Figure 3 reveals that the $\beta$ non-Gaussianity on a small
scale at high redshift are about equal to that on a large scale at
low redshift. This result indicates that the turbulence state on
scales 0.1 - 1 h$^{-1}$Mpc at redshift $z\sim 5$ would be the same
as that on scales 1 - 10 h$^{-1}$Mpc at redshift $\sim 2$. This is
consist with the fact that the Jeans length at $z\sim 5$ is less
than 0.1 h$^{-1}$Mpc and increases to $\sim 1$ h$^{-1}$Mpc at $z\sim
2$.

Quasar's Ly$\alpha$ absorption spectra at high redshift $z\sim 5 -6$
are significantly different from that at low redshift $z\sim 2 - 3$.
The latter show Ly$\alpha$ forests, while the former consist of
complete absorption troughs (Gunn-Peterson troughs) separated by
tiny transparent regions. In other words, Ly$\alpha$ transmitted
flux experience a strong evolution with redshift rising from $z\sim
2 - 3$ to $5 - 6$. The low-order statistics of the Ly$\alpha$
transmitted flux, such as the mean optical depth and its variance,
also show strong redshift evolution. Therefore, the weak
redshift-evolution of $\beta$ is very interesting. It implies that
the $\beta$ non-Gaussian feature is mainly dependent on the
nonlinear state of the fluid, but weakly dependent on the optical
depth and its variance.

\begin{figure*}
\center
\includegraphics[scale=0.31,angle=-90]{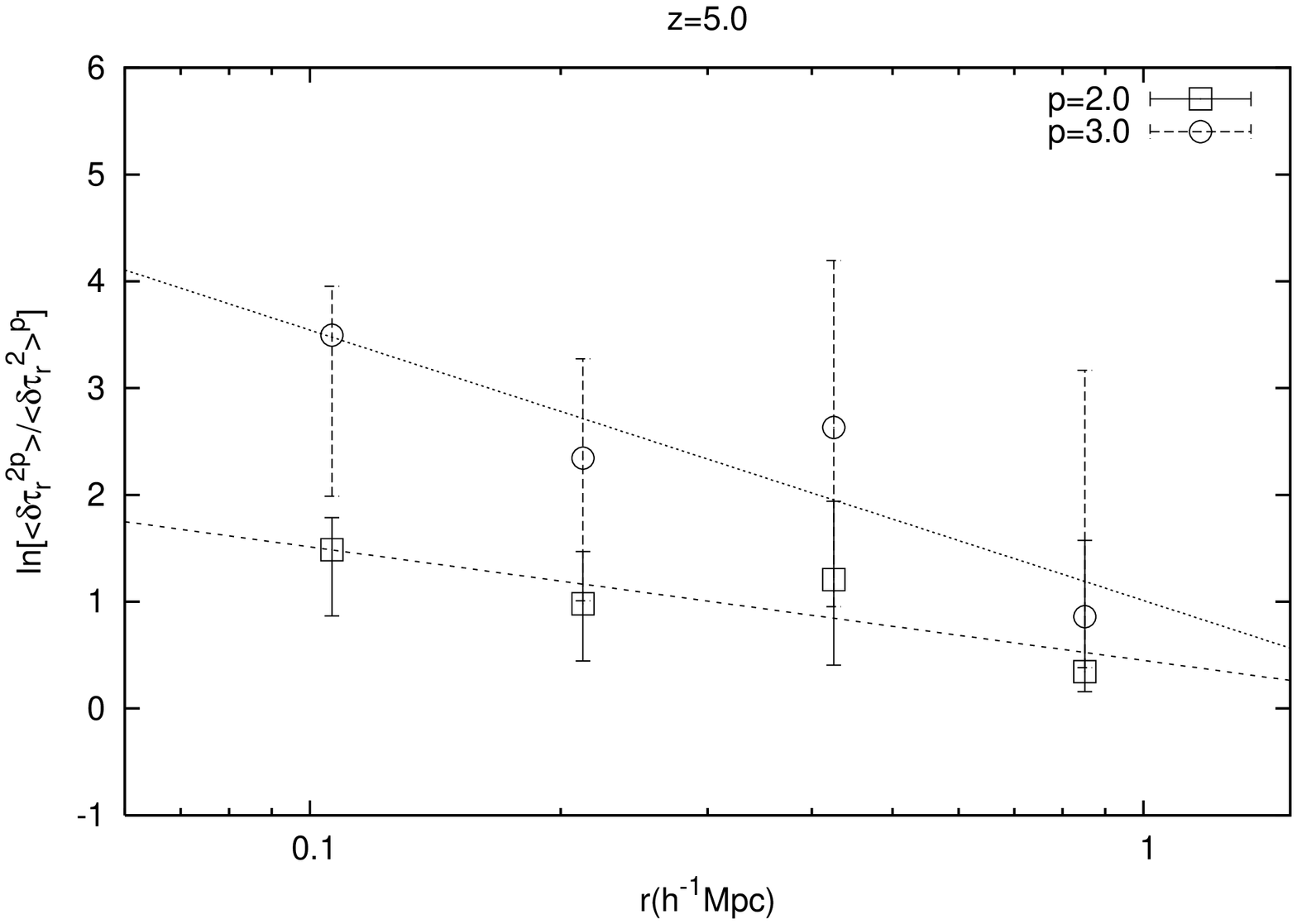}
\includegraphics[scale=0.31,angle=-90]{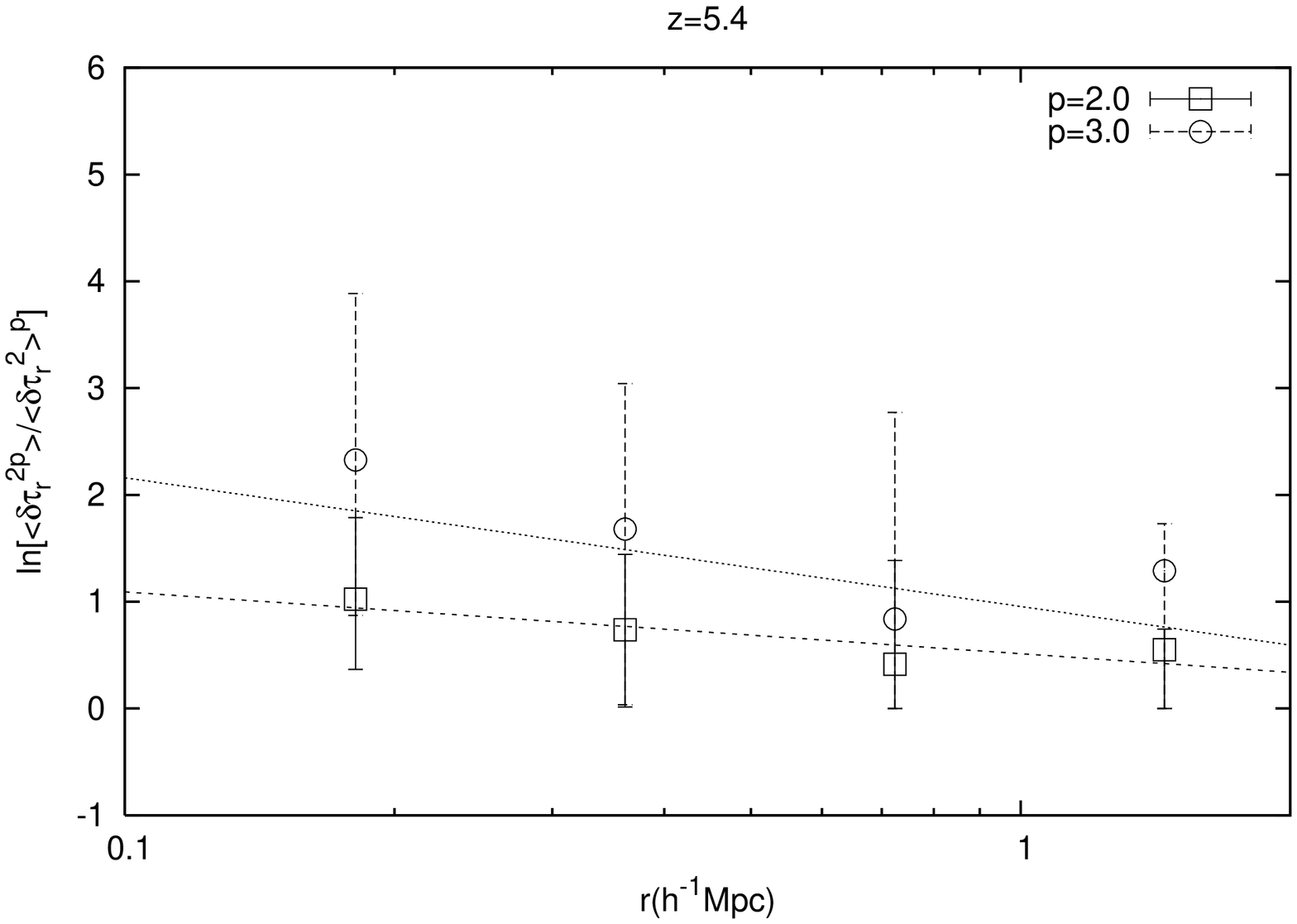}
\includegraphics[scale=0.31,angle=-90]{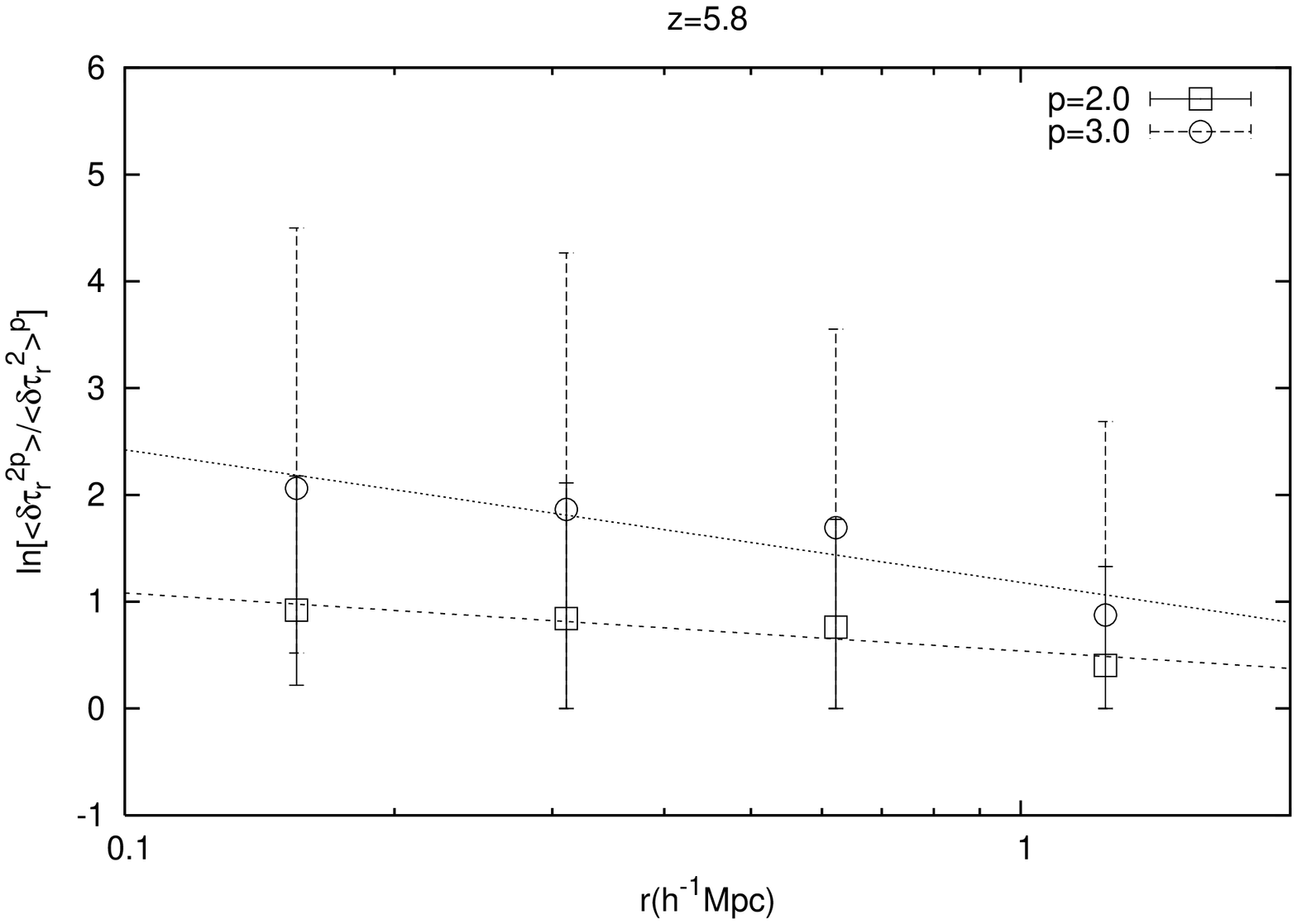}
\includegraphics[scale=0.31,angle=-90]{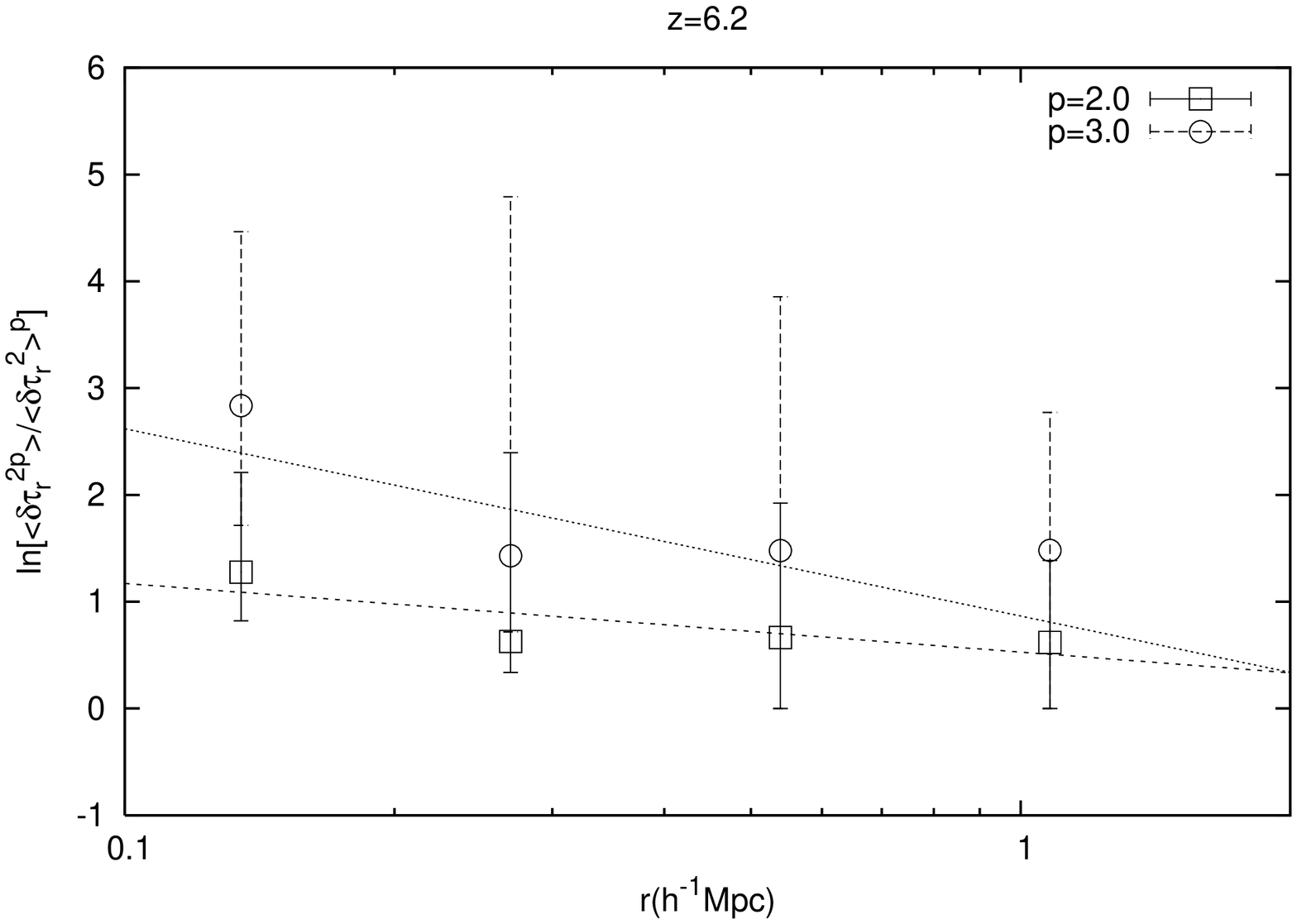}
\caption{$\ln
[\langle(\delta\tau_r)^{2p}\rangle/\langle(\delta\tau_r)^{2}\rangle^p]$
vs. $r$ of observed sample of the Ly$\alpha$
transmitted flux for redshifts at 5.0 (top-left),
5.4(top-right), 5.8(bottom-left), and 6.2(bottom-right). $p$ is
taken to be 2 (bottom line) and 3 (top line). The solid lines are
given by the least square fitting. The error bars are given by the
maximum and minimum of bootstrap re-sampling.}
\end{figure*}

\subsection{Non-Gaussianity related to parameter $\gamma$}

We now turn to the $\gamma$-related non-Gaussianity, which is given by
(Liu \& Fang 2008)
\begin{equation}
\ln \frac{\langle \delta\tau_r^{2p}\rangle}{\langle
\delta\tau_r^{2}\rangle^p}= K_p\ln r +{\rm const},
\end{equation}
and
\begin{equation}
K_p=-\gamma\frac{p(1-\beta^2)-(1-\beta^{2p})}{1-\beta}.
\end{equation}
It requires that for a given $p$, the relation of
$\ln\langle(\delta\tau_r)^{2p}\rangle/\langle(\delta\tau_r)^{2}\rangle^p$
and $\ln r$ has to be a straight line with slope $K_p$. Since
parameter $\beta$ is already determined by $\beta$-hierarchy in last
section, we can then figure out the parameter $\gamma$ from $K_p$.

\begin{table}
\center{Table 1. Parameter $\gamma$ at redshift $z=5.0$ to 6.2}
\vspace{-5mm}
\begin{center}
\bigskip
  \begin{tabular}{l|ccc}
    \hline
    z           & $p$    & $K_p$   & $\gamma$     \\
        \hline
    5.0 & 2  & -0.46$\pm$ 0.20 & $0.41^{+0.17}_{-0.18}$\\
       &  3 & -1.10$\pm$ 0.40 &  $0.43^{+0.16}_{-0.15}$   \\
    5.4 & 2  & -0.25$\pm$ 0.10 & $0.22^{+0.08}_{-0.07}$\\
       &  3 & -0.52$\pm$ 0.30 &  $0.20^{+0.12}_{-0.11}$   \\
    5.8 & 2  & -0.24$\pm$ 0.07 & $0.21^{+0.06}_{-0.06}$\\
       &  3 & -0.52$\pm$ 0.15 &  $0.21^{+0.06}_{-0.06}$   \\
    6.2 & 2  & -0.28$\pm$ 0.16 & $0.27^{+0.15}_{-0.15}$\\
       &  3 & -0.78$\pm$ 0.35 &  $0.31^{+0.13}_{-0.12}$   \\
     \hline
\end{tabular}
\end{center}
\end{table}

The result is presented in Figure 4. It shows the relation of $\ln
(\langle \delta\tau_r^{2p}\rangle/\langle
(\delta\tau_r)^{2}\rangle^p)$ vs. $\ln r$ for observational data at
redshift ranges $z=5.0$, 5.4, 5.8 and 6.2. The order parameter $p$
is set to  2 and 3, i.e. the statistics of eqs.(14) and (15) are of
the order of 4 and 6. The error bars are given by the maximum and
minimum of each $r$. The $\ln r$-dependency of $\ln (\langle
\delta\tau_r^{2p}\rangle/\langle (\delta\tau_r)^{2}\rangle^p)$
approximately at each range can be given by a straight line.

The slopes $K_p$ of the straight lines of Figure 4 are listed in
Table 1, and parameter $\gamma$ given by eq.(15) is also listed. In
log-Poisson hierarchy, parameter $\gamma$ has to be independent of
$p$ and therefore, the values of $\gamma$ determined by $K_p$ with
different straight lines should be the same. Table 1 indeed confirms
this point. We see that for a given redshift, the statistics of
$p=2$ and 3 yield the same $\gamma$ within their errors. Therefore,
the high redshift Ly$\alpha$ transmitted flux well fulfills the
$\gamma$-related non-Gaussianity.

A basic feature of log-Poisson model is to yield nonlinear terms of
$p$, i.e. the term $\beta^p$ in eq.(7) and $\beta^{2p}$ in eq.(15)
(Frisch 1995). The terms with linear $p$ can also be given by other
models. Since $\beta<1$, the tests with $\beta^n$ and $n>6$ don't
give new test on the log-Poisson model. Therefore, the statistical
order $n$  generally is taken to be less than 6.

\section[]{Hydrodynamic simulation samples}

Although the mean optical depth and its variance of Ly$\alpha$
transmitted flux of quasar's absorption spectrum underwent a strong
evolution at high redshift, the log-Poisson non-Gaussianity, as last
section reveals, shows only weak dependence on redshift. Thus, an
important question is whether the two aspects of the redshift
evolutions can be conciliated within the concordance $\Lambda$CDM
model. We study this problem with cosmological hydrodynamic
simulation samples.

\subsection{Simulation samples of neutral hydrogen}

We first produce the samples of mass density, temperature and
velocity fields of cosmic hydrogen with hydrodynamic simulation with
the same code of Liu et al. (2007, 2008), which is based on Eulerian
method for hydrodynamics with 5th order Weighted Essentially
Non-oscillatory (WENO) finite difference scheme and particle
mesh(PM) method for dark matter particles (Feng et al. 2004). We use
cosmological parameters given by the latest result of WMAP (Komatsu
et al. 2009). We run simulations in a period box of side $25 h^{-1}$
Mpc with $512^{3}$ grids and dark matter particles which have a mass
resolution of $1.04 \times 10^{7} M_{\odot}$. Assuming ionization
equilibrium, atomic processes, including radiative cooling, heating
and fraction of species, are modeled using the primordial
composition $(X=0.76,Y=0.24)$ and formalism in the Appendix of
Theuns et al. (1998), under optically thin approximation.
Photoionization and photoheating are switched on after the UV
background is added at $z=11.0$.

Started at $z=99$, a sample output at $z=11$ first. Simulations with
different UV histories are then performed from this snapshot at
$z=11$ and produce snapshots at redshifts from $z=6.5$ to $z=4.9$ at
a interval of $\Delta z =0.1$. As we focus on the evolution at high
redshifts, simulations are stopped at $z=4$. From snapshot dumps, we
produce the mass density field of hydrogen $\rho({\bf x})$,
temperature field $T({\bf x)}$, and velocity field ${\bf v}({\bf
x})$. Star formation is not included. However, the contribution of
stars to the UV background is considered by fitting the redshift
evolution of the UV background with mean optical depth and its
variance of Ly$\alpha$ transmitted flux field. This code has
recently been used to produce samples to show the turbulence
behavior at redshift as high as $z\simeq 4$ (Zhu et al. 2010). These
samples would also be suitable to test the log-Poisson non-Gaussianity of
the IGM at high redshifts.

Although the Gunn-Peterson optical depth shows dramatic decrease
with redshift and abnormally large scatter at $z \sim 6$, it can
still be fitted by models of a uniform ionizing background (Lidz et
al. 2006; Liu et al. 2006, 2007; Mesinger \& Furlanetto 2009). As
current observation does not give a well knowledge of ionizing source
at high redshift, instead of the Haardt \& Madau (2001) model of UV
background history, we use a more general uniform UV background with
the ionizing photons have a power-law spectrum with index $-1.0$ and
an normalized coefficient $J_{21}$. The hydrogen photoionization
rate $\Gamma_{-12}$ is calculated using the fitting formula in
Theuns et al. (1998) and then is used to calculating the heating and
cooling in the hydrodynamic simulation. The photoionization rate can
be given by
\begin{equation}
\Gamma_{-12}=3.15 J_{21}.
\end{equation}

\begin{figure*}
\center
\includegraphics[scale=0.32,angle=-90]{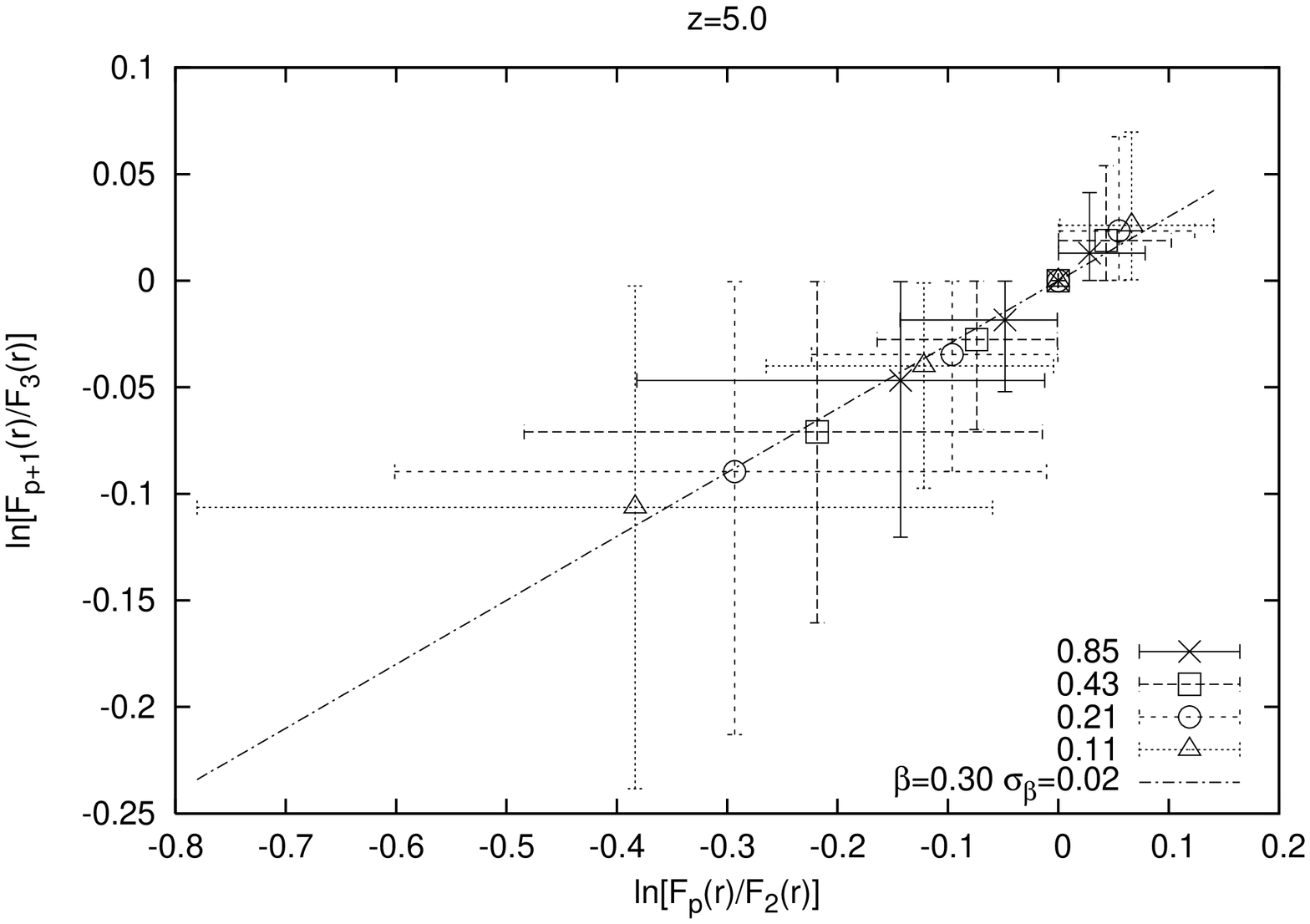}
\includegraphics[scale=0.32,angle=-90]{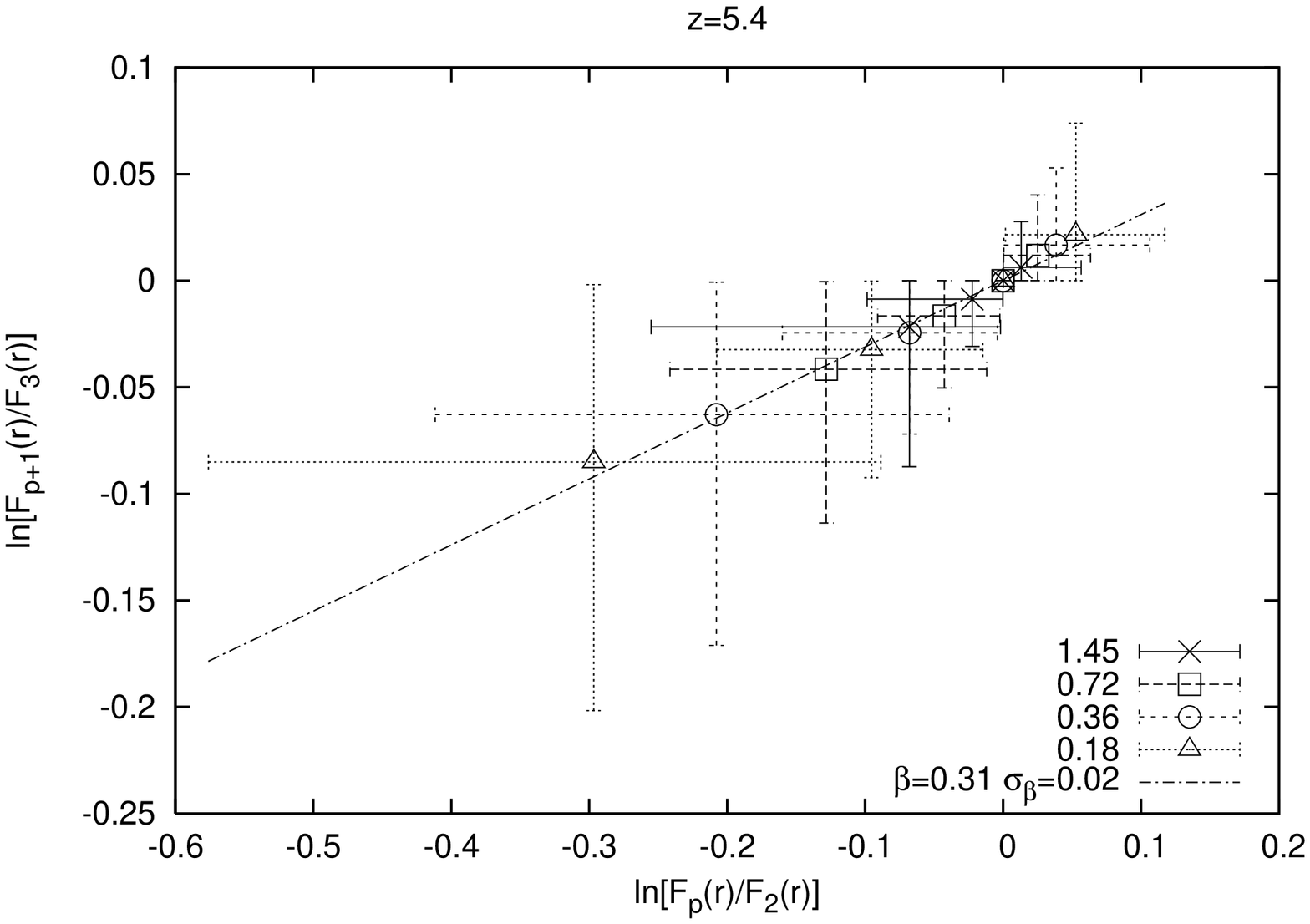}
\includegraphics[scale=0.32,angle=-90]{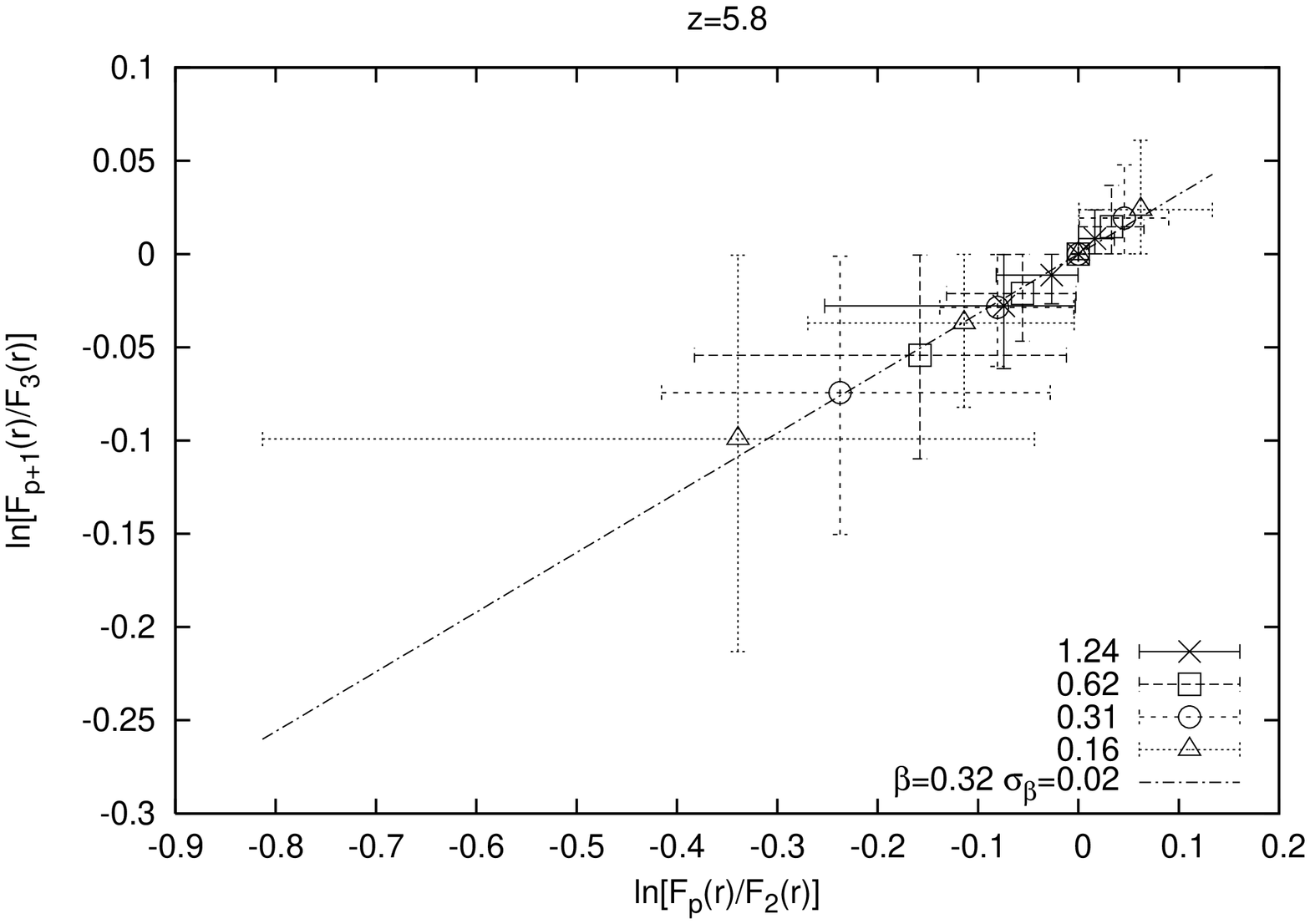}
\includegraphics[scale=0.32,angle=-90]{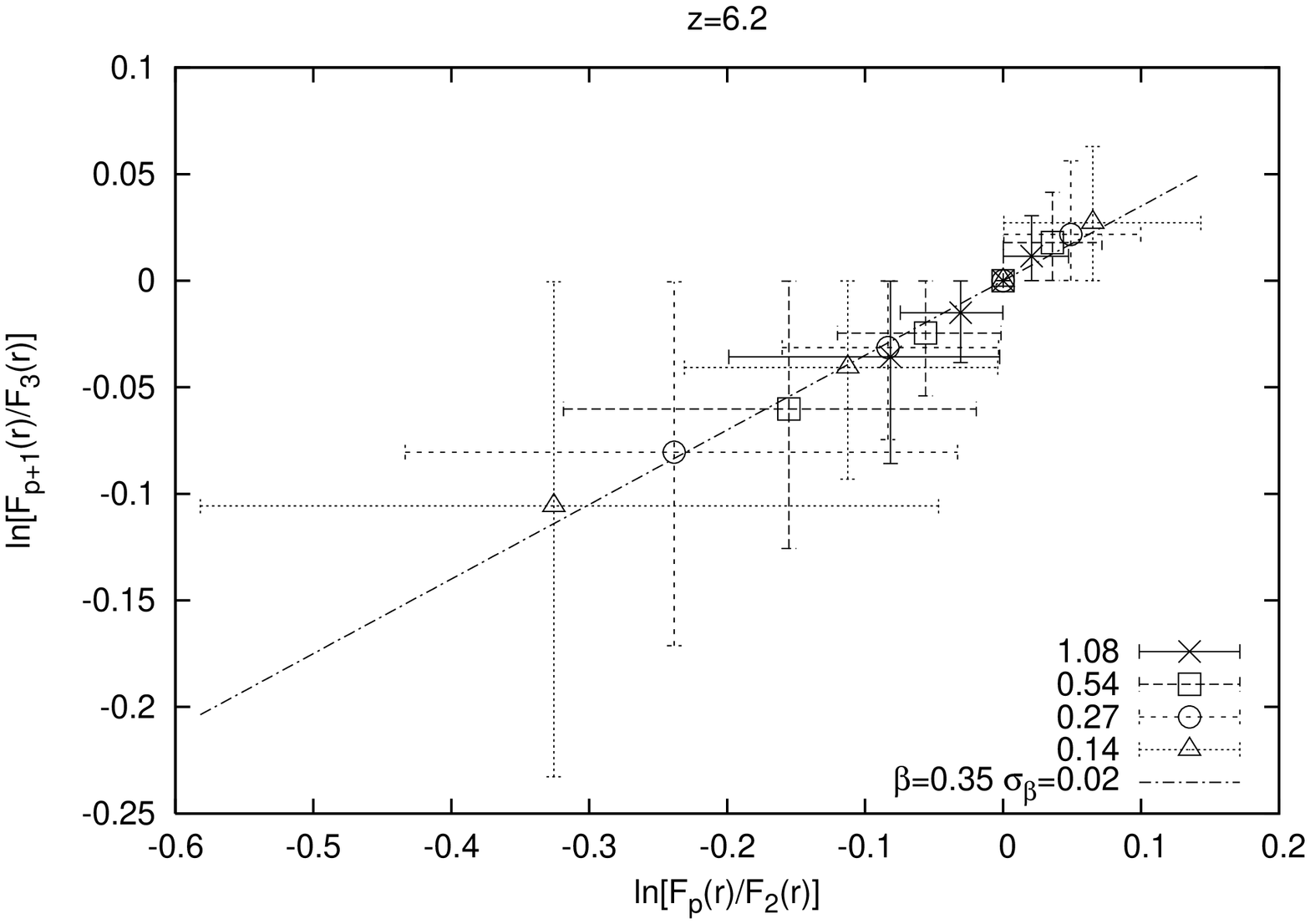}
\caption{$\ln [F_{p+1}(r)/F_3(r)]$ vs. $\ln [F_p(r)/F_2(r)]$ for
simulation samples of the mass density field of neutral hydrogen
at redshifts 5.0, 5.4, 5.8, 6.2. The statistical order to be $p=$
1, 1.5, 2, and 2.5.}
\end{figure*}
\begin{figure*}
\center
\includegraphics[scale=0.30,angle=-90]{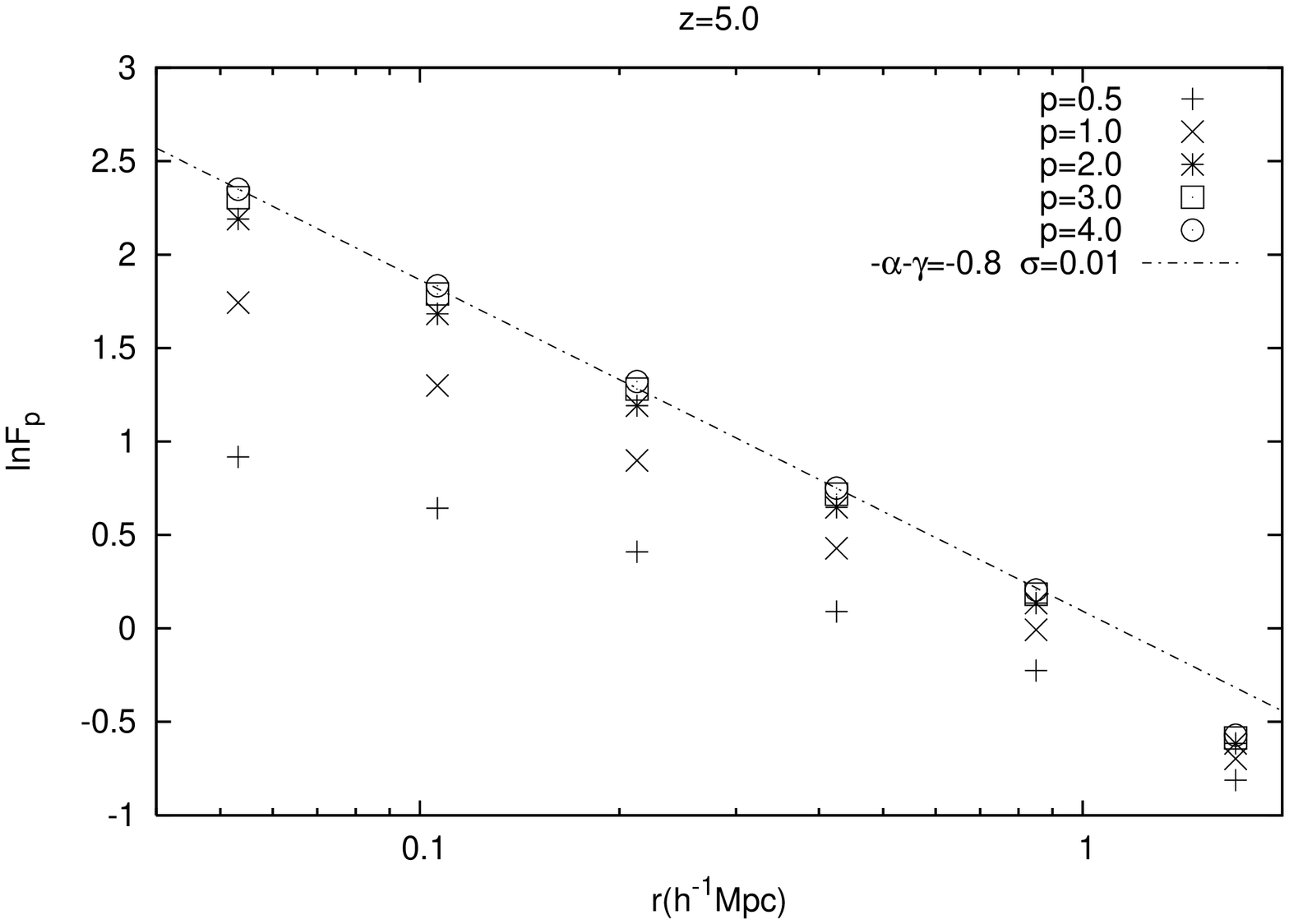}
\includegraphics[scale=0.30,angle=-90]{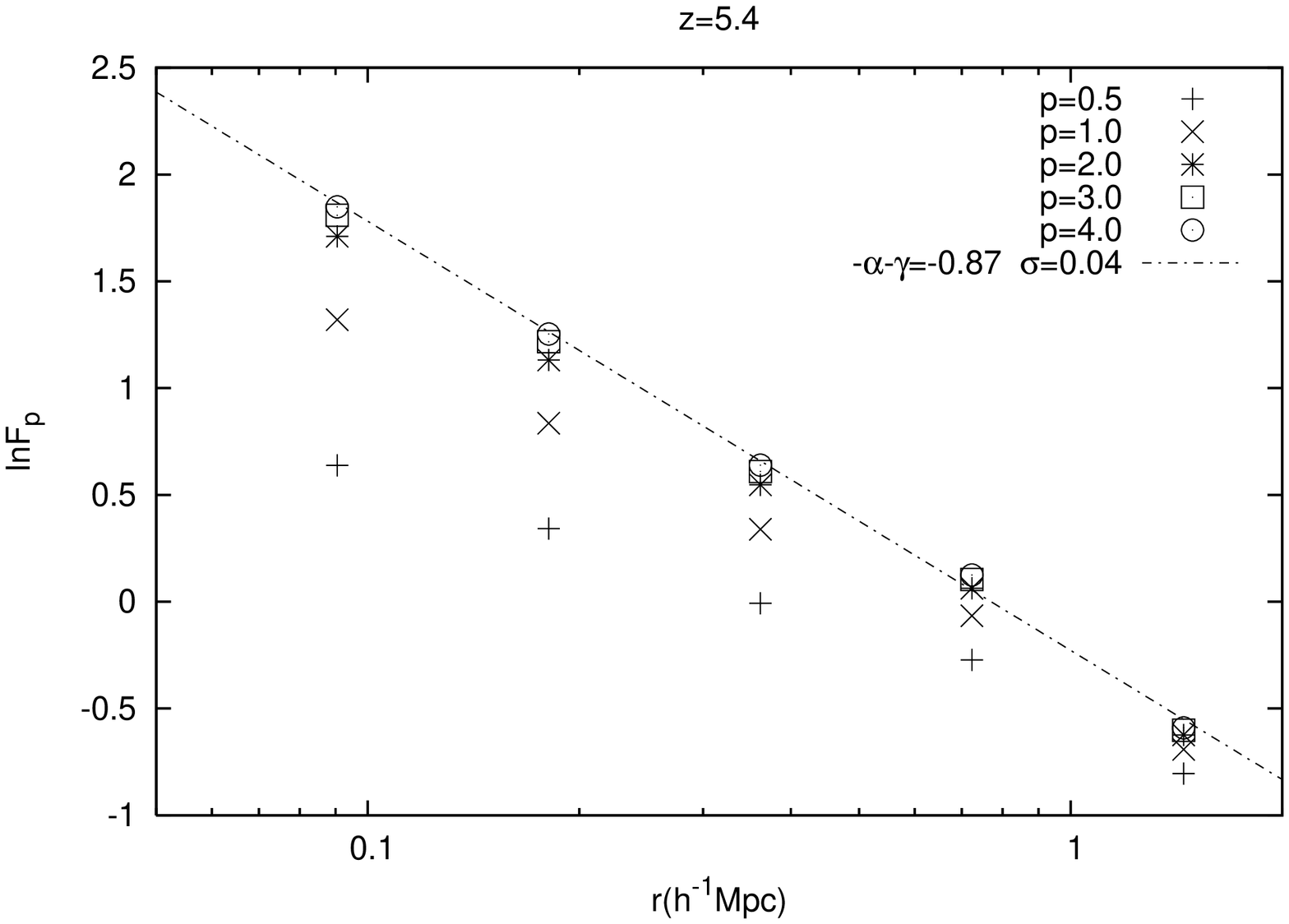}
\includegraphics[scale=0.30,angle=-90]{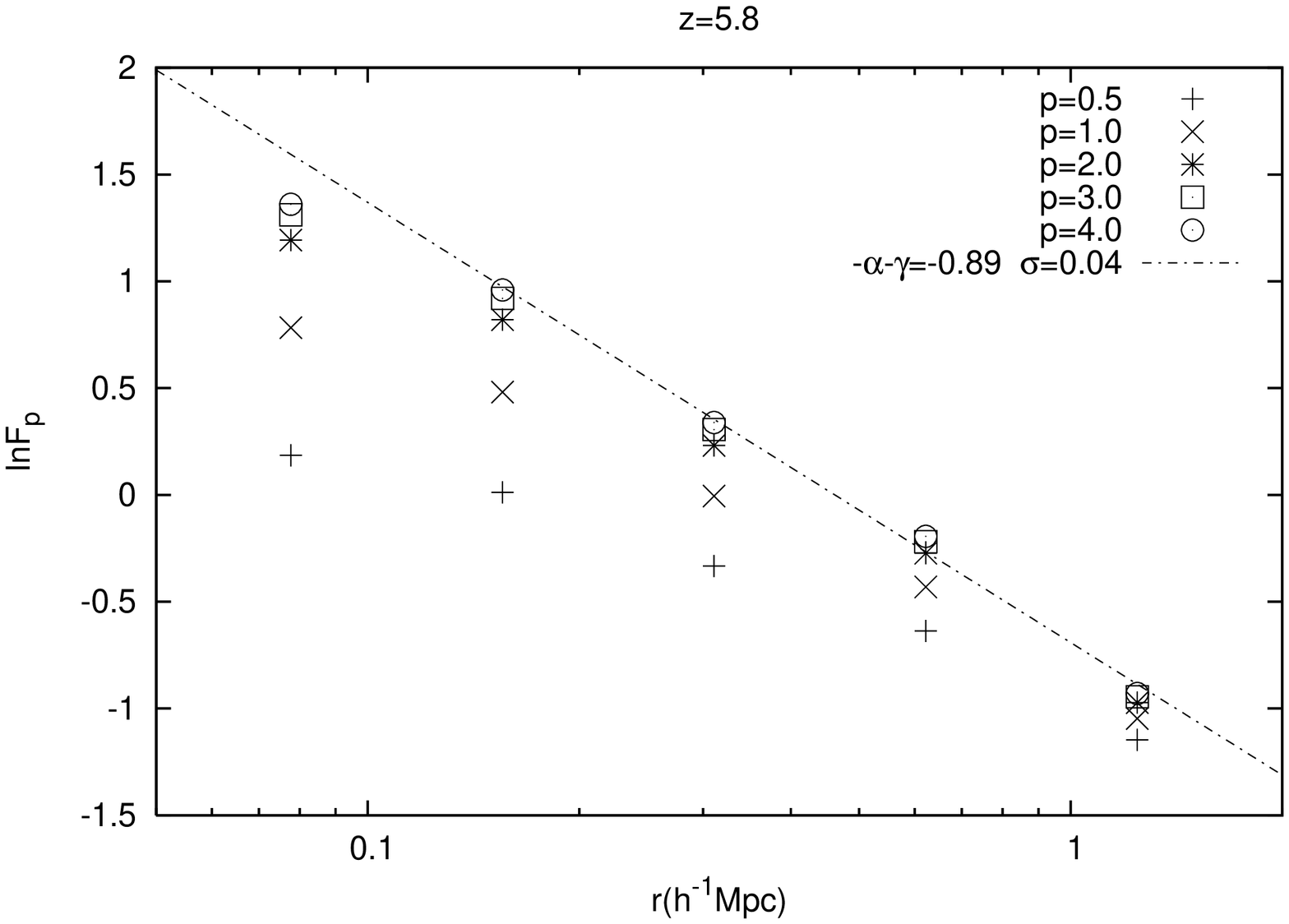}
\includegraphics[scale=0.30,angle=-90]{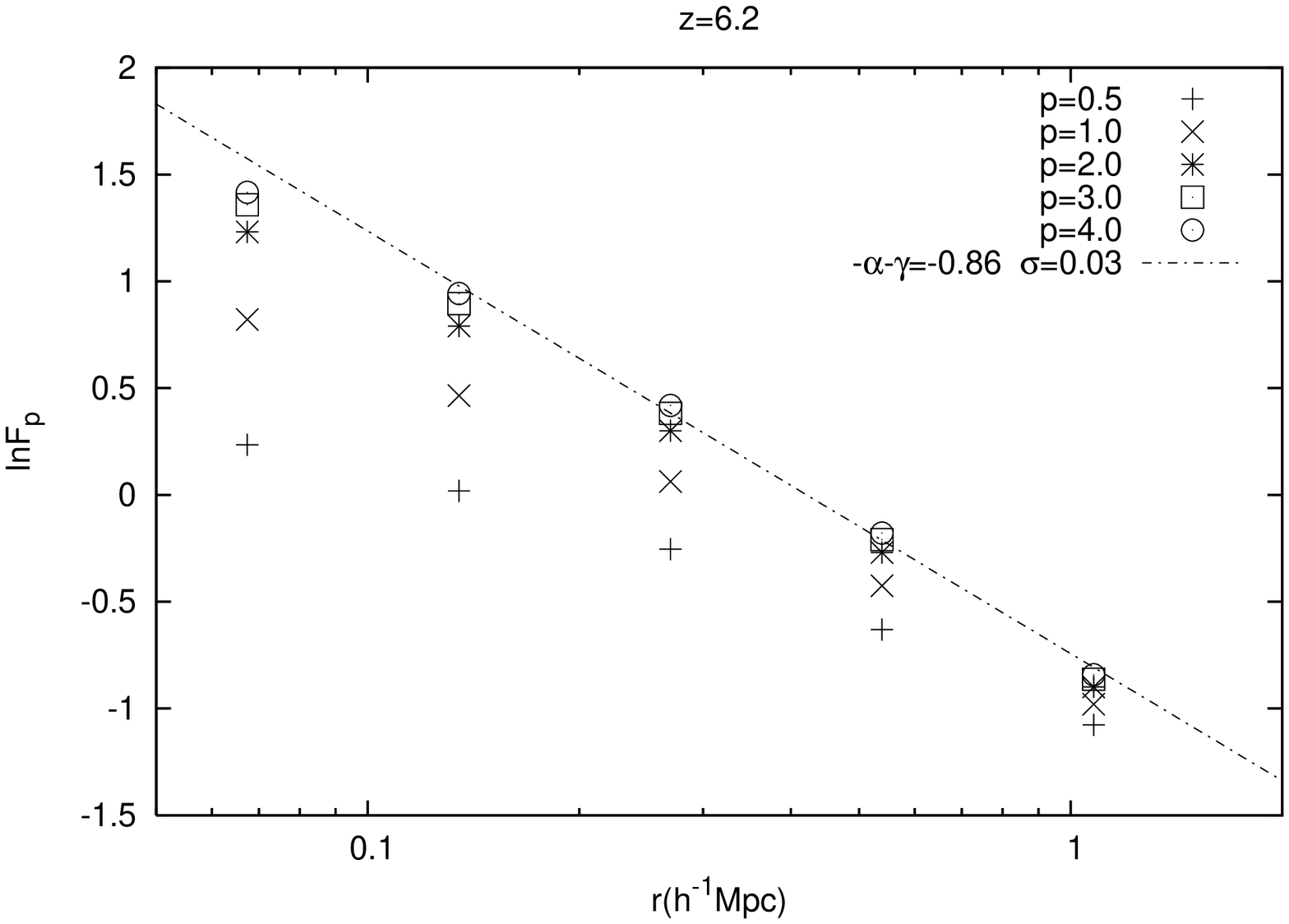}
\caption{$\ln F_p(r)$ vs. $\ln r$ of simulation samples of neutral
hydrogen mass density field. The number $\gamma$ is given in each panel.}
\end{figure*}

Based on the above assumption, the history of the UV background in
our simulation is given by the evolution of $J_{21}$. There is no
direct observation on this parameter. Considering the two typical
reionization history in references, one is the extended scenario and
the other is phase transition, we use the following two
redshift-dependent models of $J_{21}$,
\begin{equation}
J_{21}(z)=5.0\times \exp(-0.475z).
\end{equation}
\begin{equation}
J_{21}(z)=\exp(-0.21z^{2}+1.5z-3.0)+0.02\exp(-7.0/z).
\end{equation}
The $J_{21}(z)$ of eq.(17) can be considered as a model of the
extended reionization scenario. Eq(18) gives approximately the same
evolution as eq.(17) at $z=4.0-5.0$, but drops about an order of
magnitude from $z=5.0$ to $z=7.0$ and stays at very low level when
$z>7$. It is to mimic the reionization as a phase transition over
redshift range from $z=5.0$ to $z=7.0$. The parameters used in
eqs.(17) and (18) actually are determined by the fitting of
simulated samples with observed mean optical depth and its variance
of Ly$\alpha$ transmitted flux field. Several simulations have been performed
before these parameters are selected. It is interesting to see that
with the parameters of eqs.(17) and (18), the intensities of
$\Gamma_{-12}$ given by eq.(16) in the redshift range $z=4-9$ are
just in between of the values given by 1.) Haardt \& Madau, (2001)
and 2.) proximity effect of quasars (Dall'Aglio et al. 2008, Gilmore
et al. 2009 and references therein ).

The last but not least, although the intensities of the UV background 
given by models (17) and (18) are very different at $z=5.0$ to $z=7.0$,
the non-Gaussian features of the transmitted field of Ly$\alpha$ are less 
affected. This is because  the basic variable $\delta\tau_r=\tau(x+r)-\tau(x)$
is not sensitive to the change of the uniform ionizing background given by
the models (17) and (18), especially when $r$ is small.

\subsection{Intermittence of neutral hydrogen density field}

\begin{figure*}
\center
\includegraphics[scale=0.30,angle=-90]{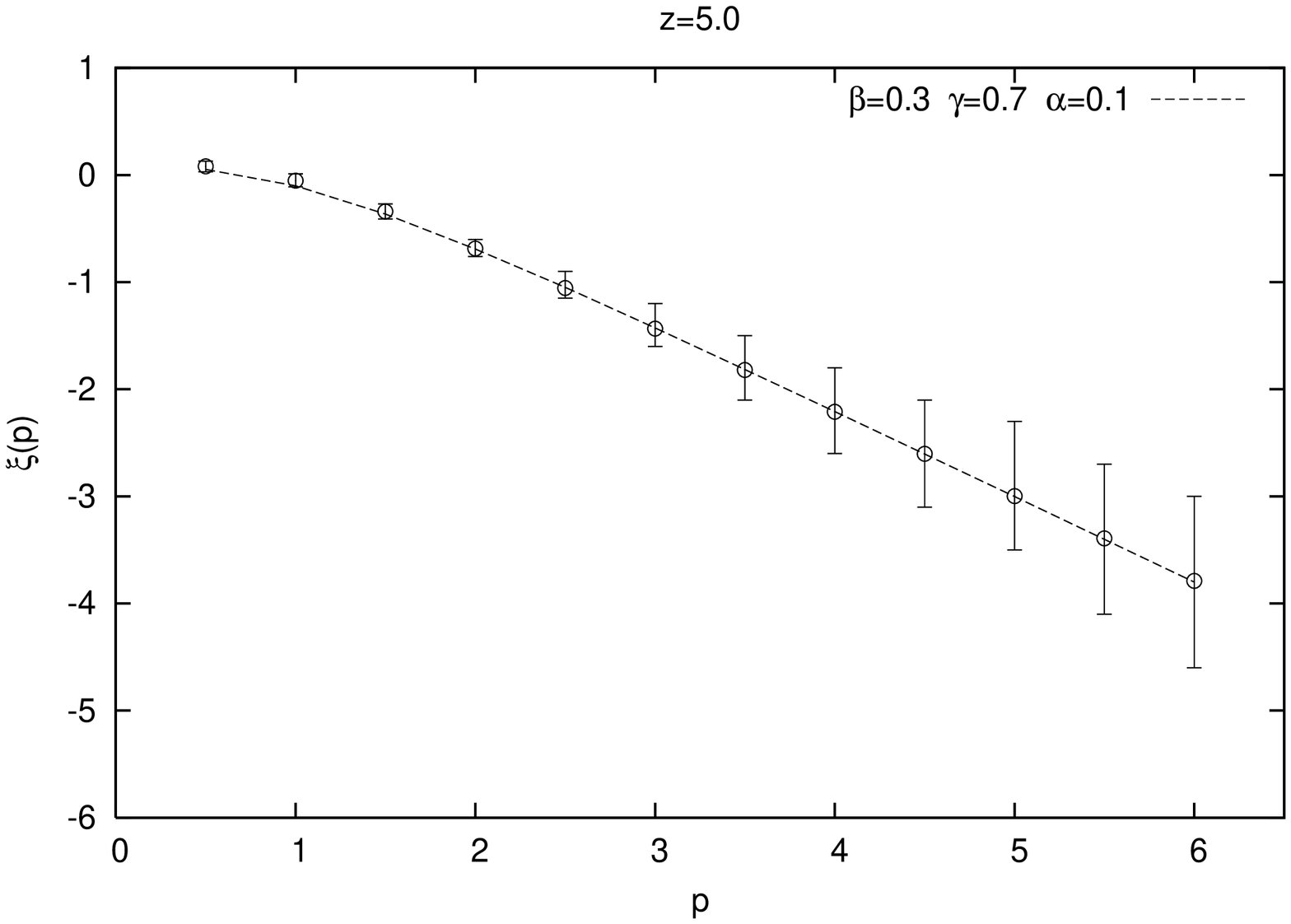}
\includegraphics[scale=0.30,angle=-90]{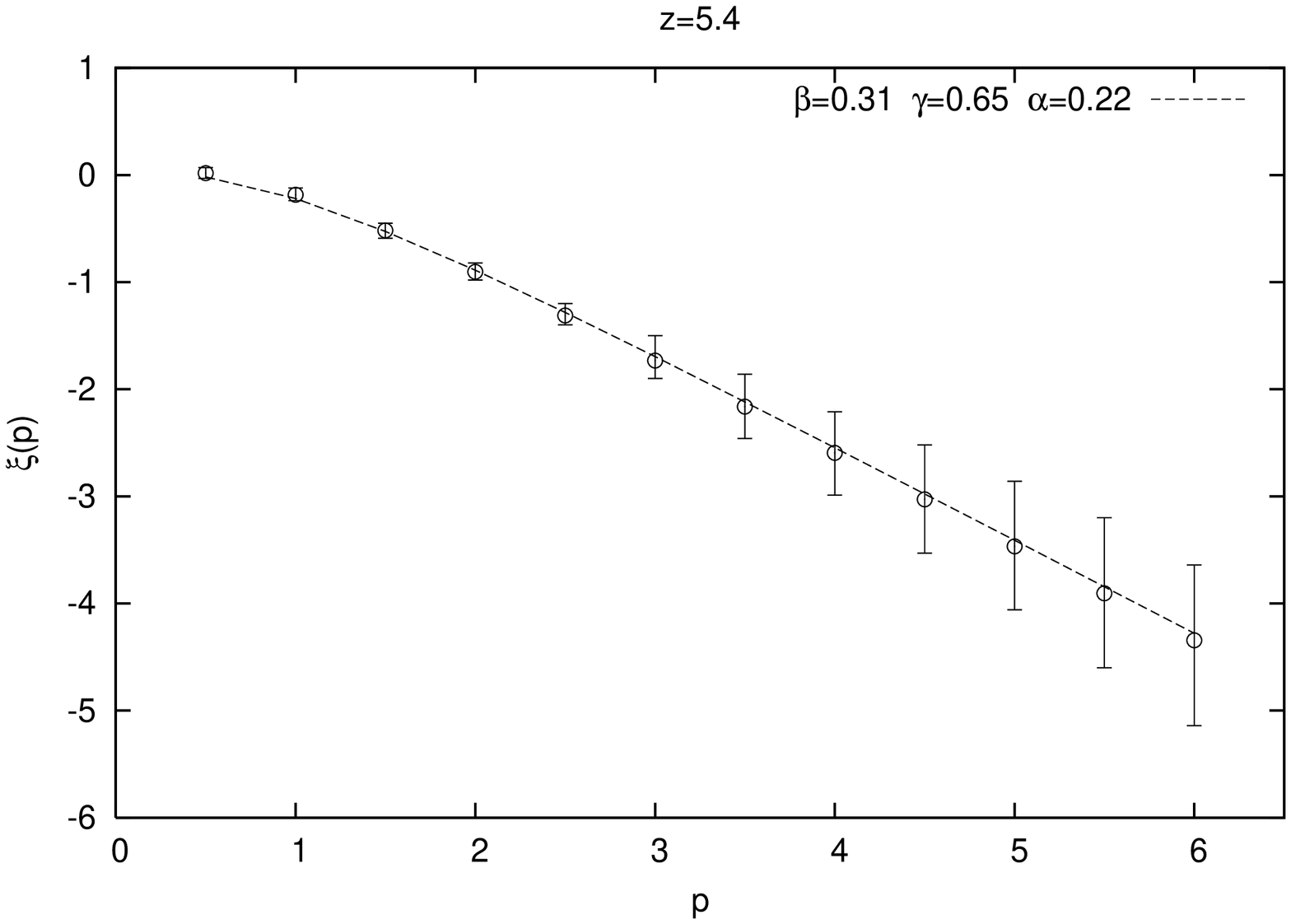}
\includegraphics[scale=0.30,angle=-90]{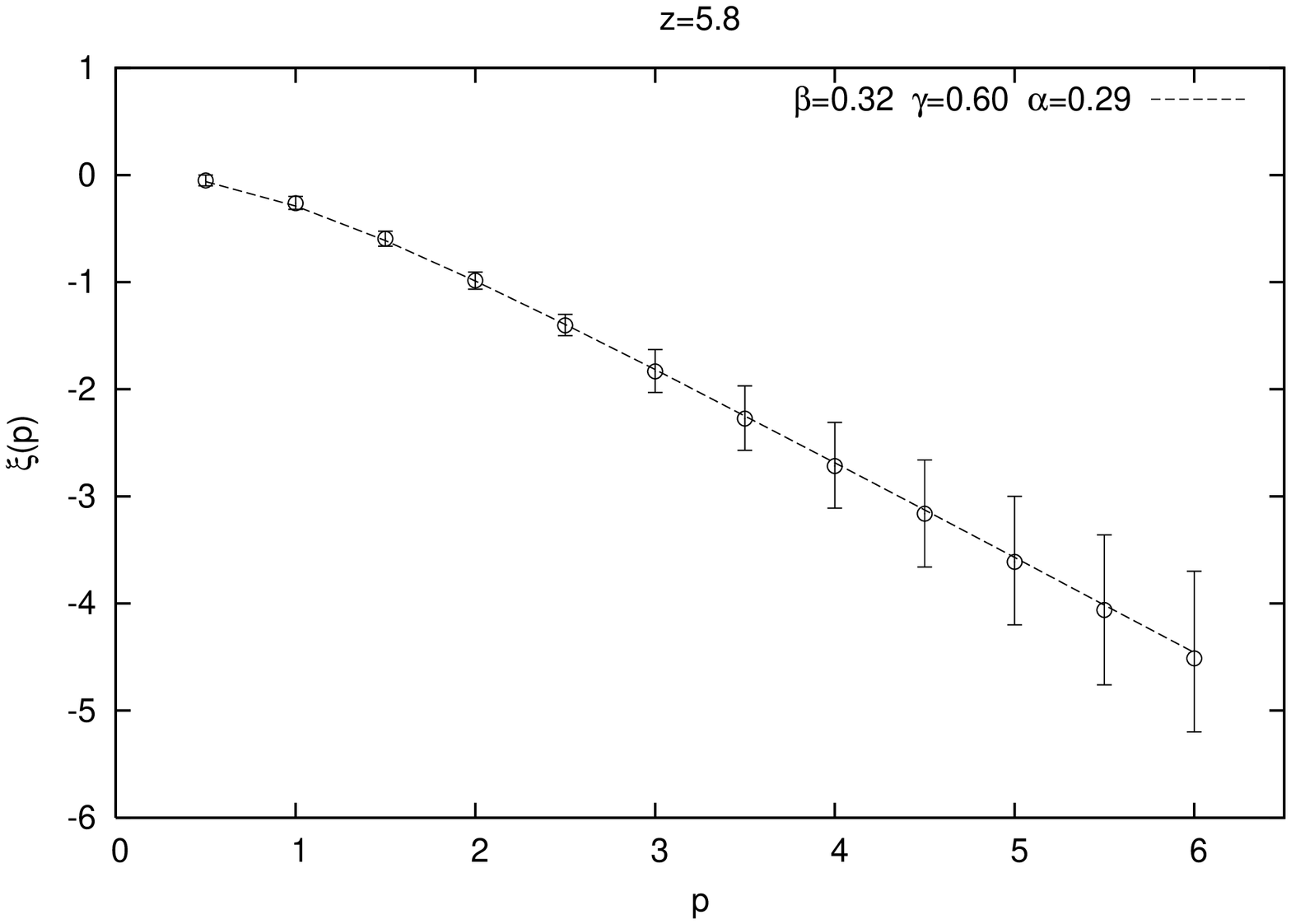}
\includegraphics[scale=0.30,angle=-90]{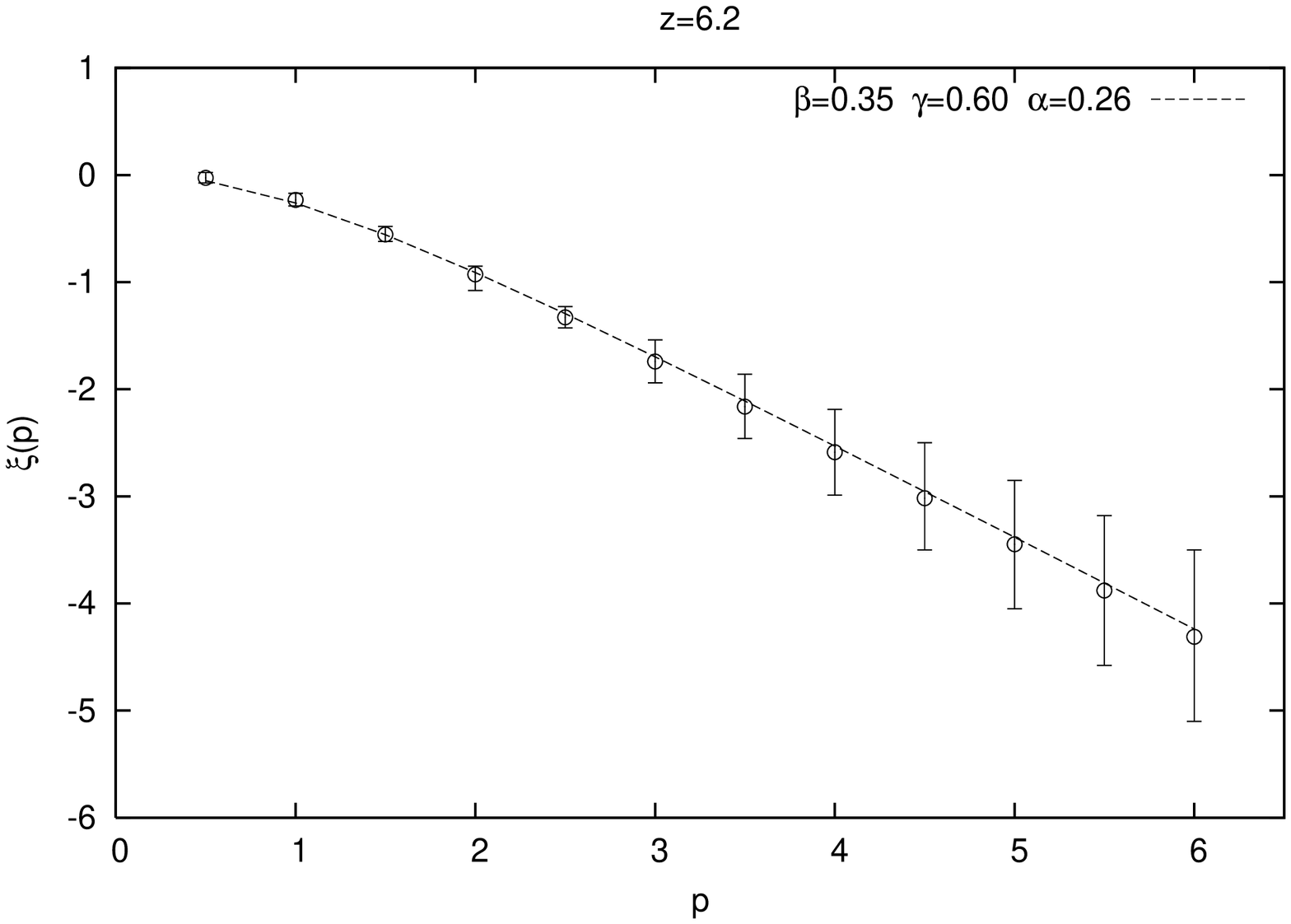}
\caption{Intermittent exponent $\xi(p)$ of simulation samples of the
mass density field of neutral hydrogen. The error bars are the
variance of $\xi(p)$ over 80 sub-samples, each of which contains of
2 one-dimensional samples.}
\end{figure*}

Before simulating the Ly$\alpha$ transmitted flux, we analyze the
log-Poisson behavior of the simulation samples of neutral hydrogen
density field, $n_{\rm HI}({\bf x})$. We first calculate the
$\beta$-hierarchy of neutral hydrogen field. The results are
presented in Figure 5, which contains all data points of
$F_{p+1}(r)/F_3(r)$ and $F_p(r)/F_2(r)$ with $p=$ 1, 1.5, 2, and
2.5, and available scale $r$. To estimate errors, we divide our 160
one-dimensional samples into 80 subsamples, each of which has 2
lines. The error bars are given by the scattering ranges of the 80
subsamples. Figure 5 shows that $\beta$-hierarchy is also well hold
for neutral hydrogen density field. The numbers of $\beta$ are also
shown in Figure 5. It is in the range $3.0 - 3.5$.

In log-Poisson hierarchy, the $r$-dependence of $F_p(r)$ is
given by (Liu et al 2008; Lu et al. 2009)
\begin{equation}
F_p(r) =A r^{-\alpha -\gamma(1-\beta^p)}.
\end{equation}
That is, the relation between $\ln F_p(r)$ and $\ln r$ should be a
straight line with the slope -$\alpha+\gamma(1-\beta^p)$. For
Gaussian field we have $\beta=1$, and therefore, the slope is
independent of $p$. For log-Poisson field ($\beta<1$), when $p$ is
large, the slope converges to $-(\alpha-\gamma)$. On the other hand,
parameter $\alpha$ can be determined by the power spectrum of
$n_{\rm HI}(x)$ (\S 2.2). Thus, The $\ln F_p(r)$-$r$ relation with
larger $p$ can be used to determine parameter $\gamma$.

Figure 6 presents $\ln F_p(r)$ vs. $\ln r$ of the simulation samples of
neutral hydrogen mass density field in the physical length scale
range of $0.1 <r< 1.5$ h$^{-1}$ Mpc and orders of $p=0.5\times n$ with
$n=1, 2...8$.  For all redshifts and $p\leq 4$, $\ln F_p(r)-\ln r$ can indeed
be approximately fitted by straight lines. These straight lines have different
slopes. The higher the $p$, the more steeper the straight lines are. It
implies that $\beta<1$ and the density fields are non-Gaussian. When $p>3$, the fitted
straight lines are almost independent of $p$. It shows the convergent of
the slope to -$(\alpha-\gamma)$ at high $p$. The parameters $\gamma$ found with
slope of $\ln F_p(r)$-$r$ lines are 0.70, 0.65, 0.60 and 0.60, corresponding to
redshifts $z=$ 5.0, 5.4, 5.8 and 6.2 respectively.

From the parameters $\beta$ and $\gamma$ given above, we plot the
intermittent exponent $\xi(p)$ [eq.(8)] in Figure 7, in which data
points are given by fitting structure functions of simulation
samples to eq.(2) with $p=0.5+0.5n$, $n=0...11$.  As mentioned in \S
3.3, the test is limited in the range $p\leq 6$.  The error bars of
Figure 7 are given by the maximum and minimum of the 80 subsamples.

One can conclude that the fields of neutral hydrogen mass density
can be well described by the She-Leveque's scaling formula with
statistical order as high as $p=6$. It should be noted again, Figure
7 covers the physical scales from $\sim 0.1$ to 1.5 h$^{-1}$Mpc. It
indicates that cosmic neutral hydrogen fluid is turbulent on the
scales considered.

\section[]{Confrontation of observed data with simulation simples}

\subsection{Ly$\alpha$ spectrum synthesize}

\begin{figure*}
\center
\includegraphics[scale=0.35,angle=-90]{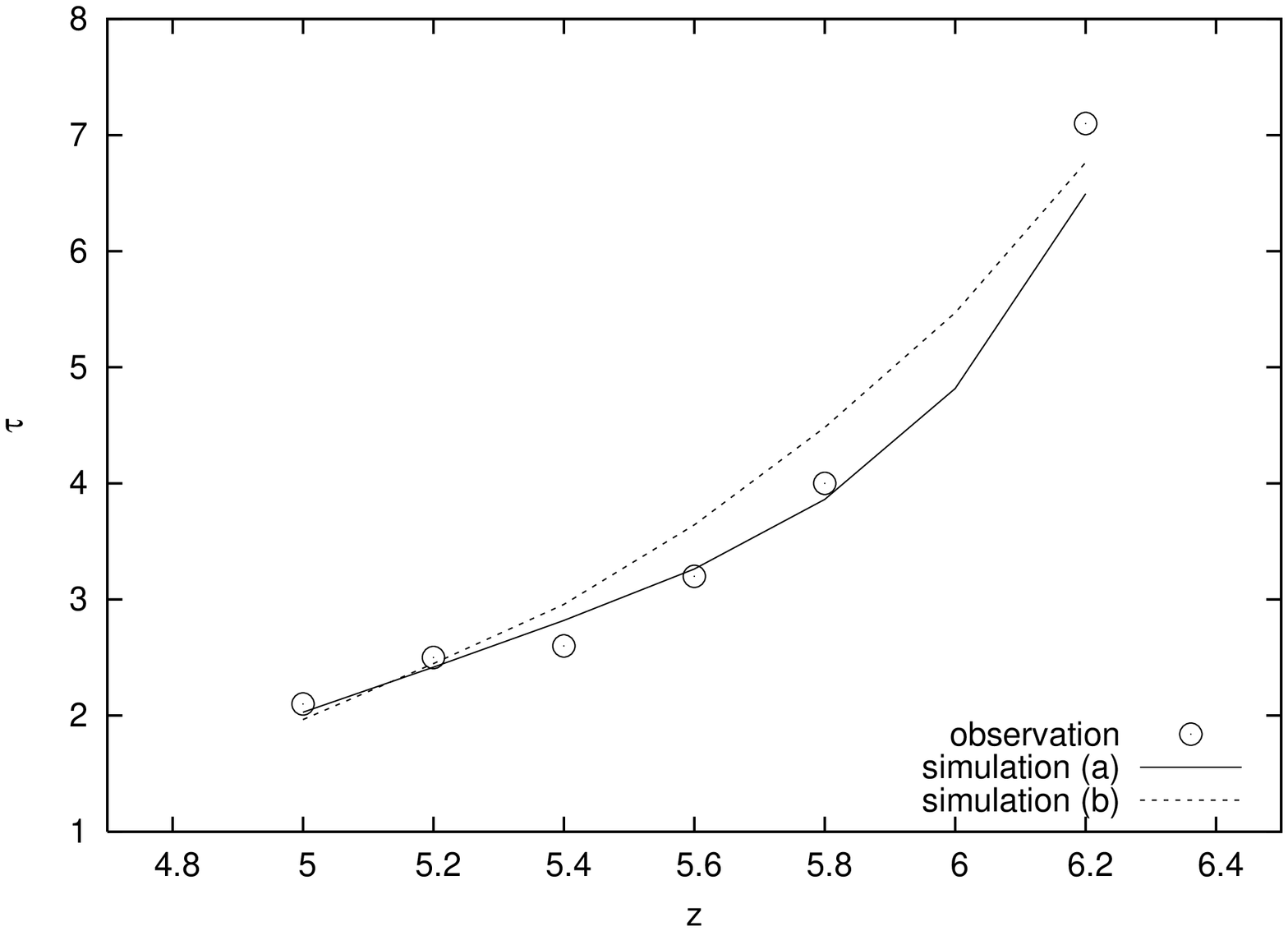}
\includegraphics[scale=0.35,angle=-90]{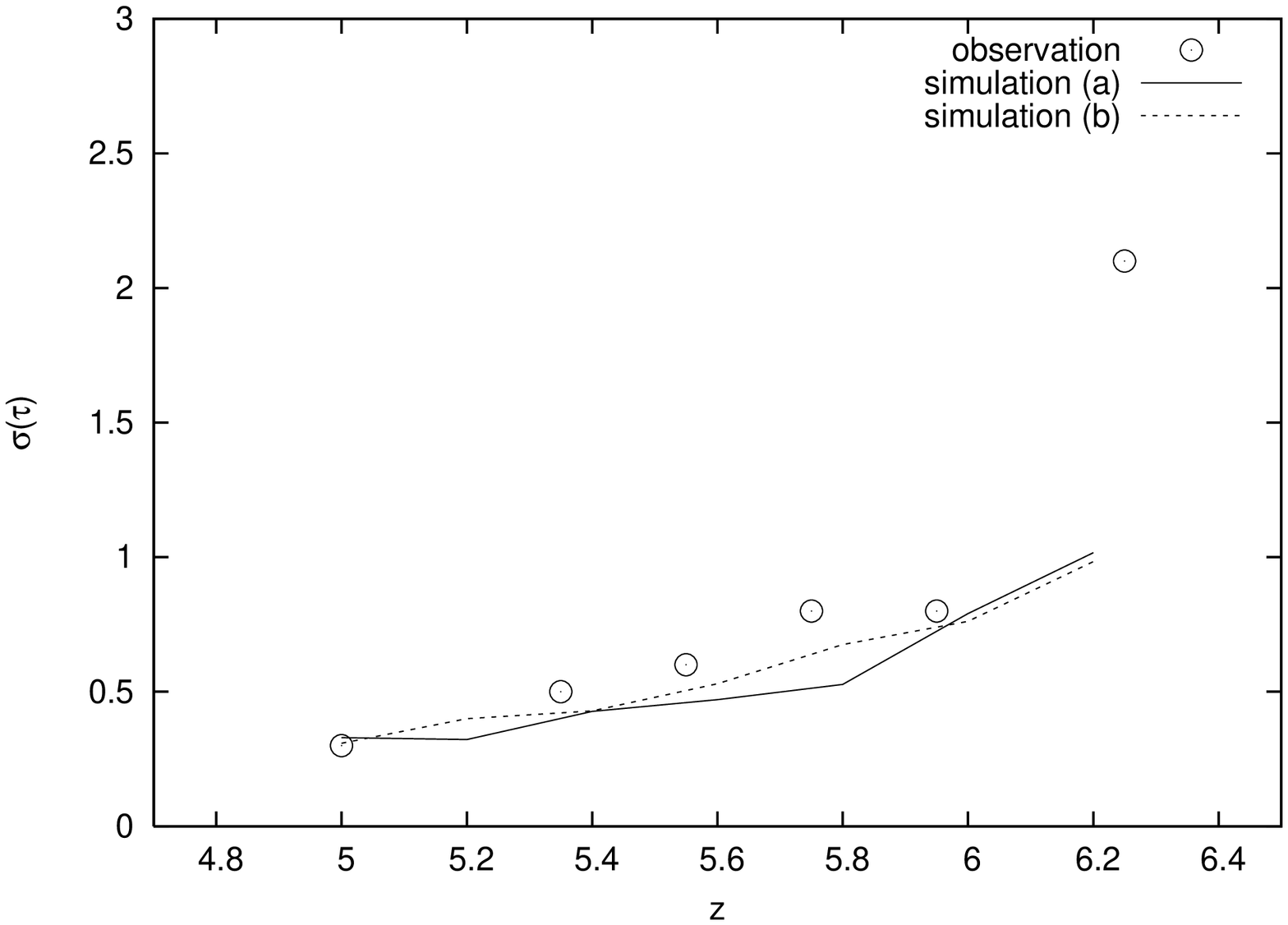}
\caption{Redshift-dependence of mean optical depth (left) and its
variance (right). It shows observed result (circle) and simulation samples
with UV background eq.(17) (solid line) and eq.(18) (dotted line).}
\end{figure*}
\begin{figure*}
\begin{center}
\includegraphics[scale=0.35]{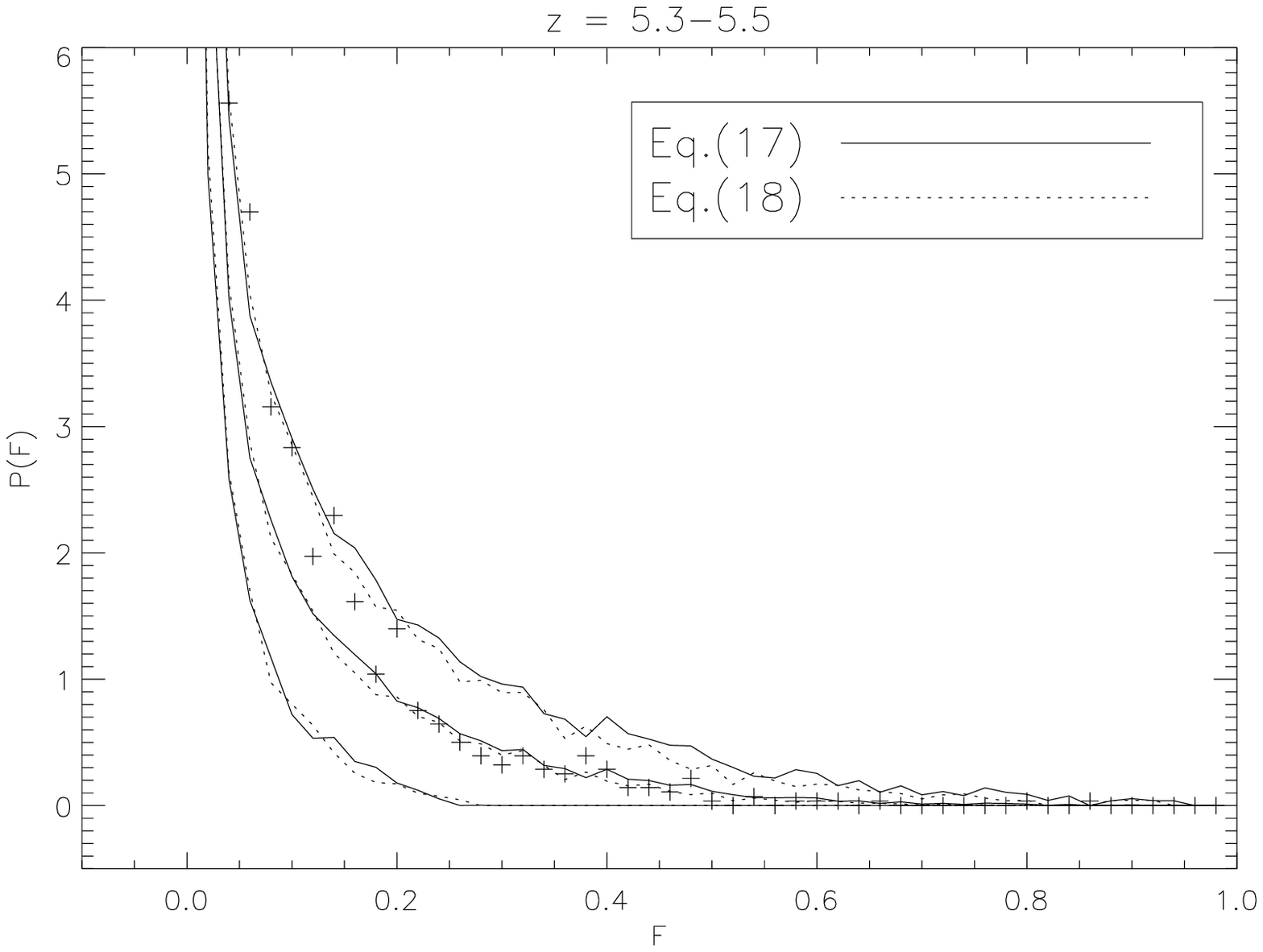}
\includegraphics[scale=0.35]{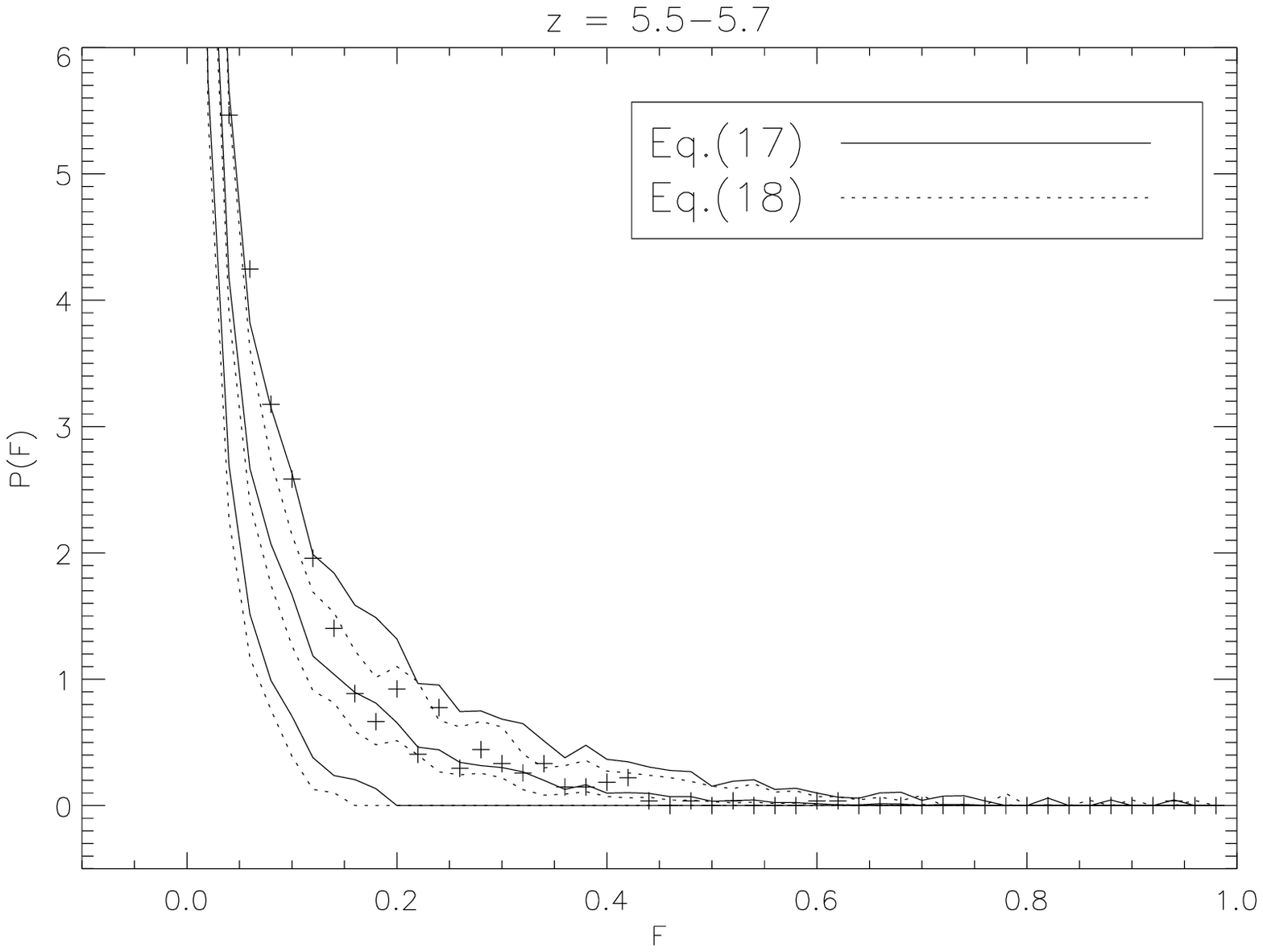}
\end{center}
\begin{center}
\includegraphics[scale=0.35]{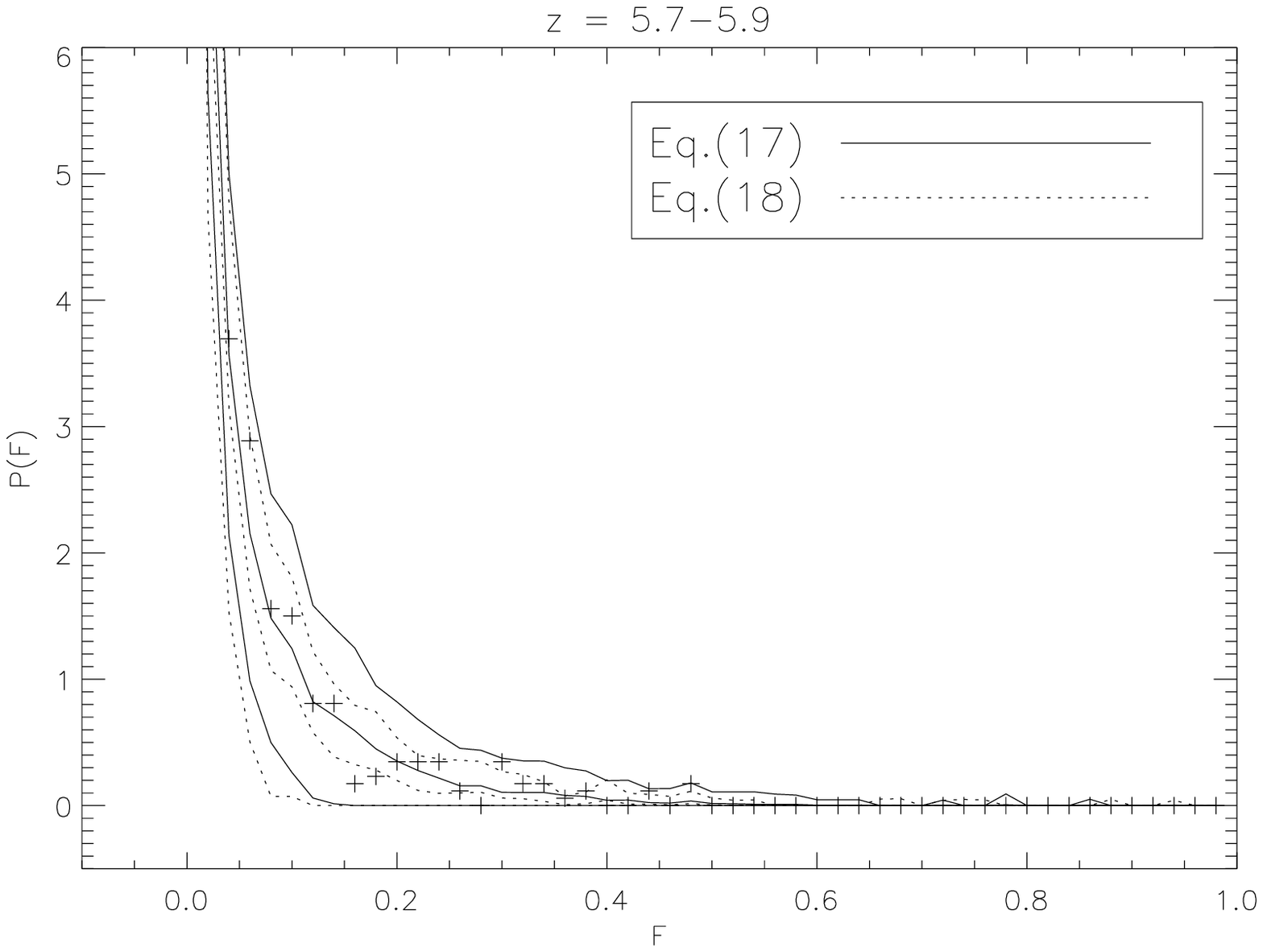}
\includegraphics[scale=0.35]{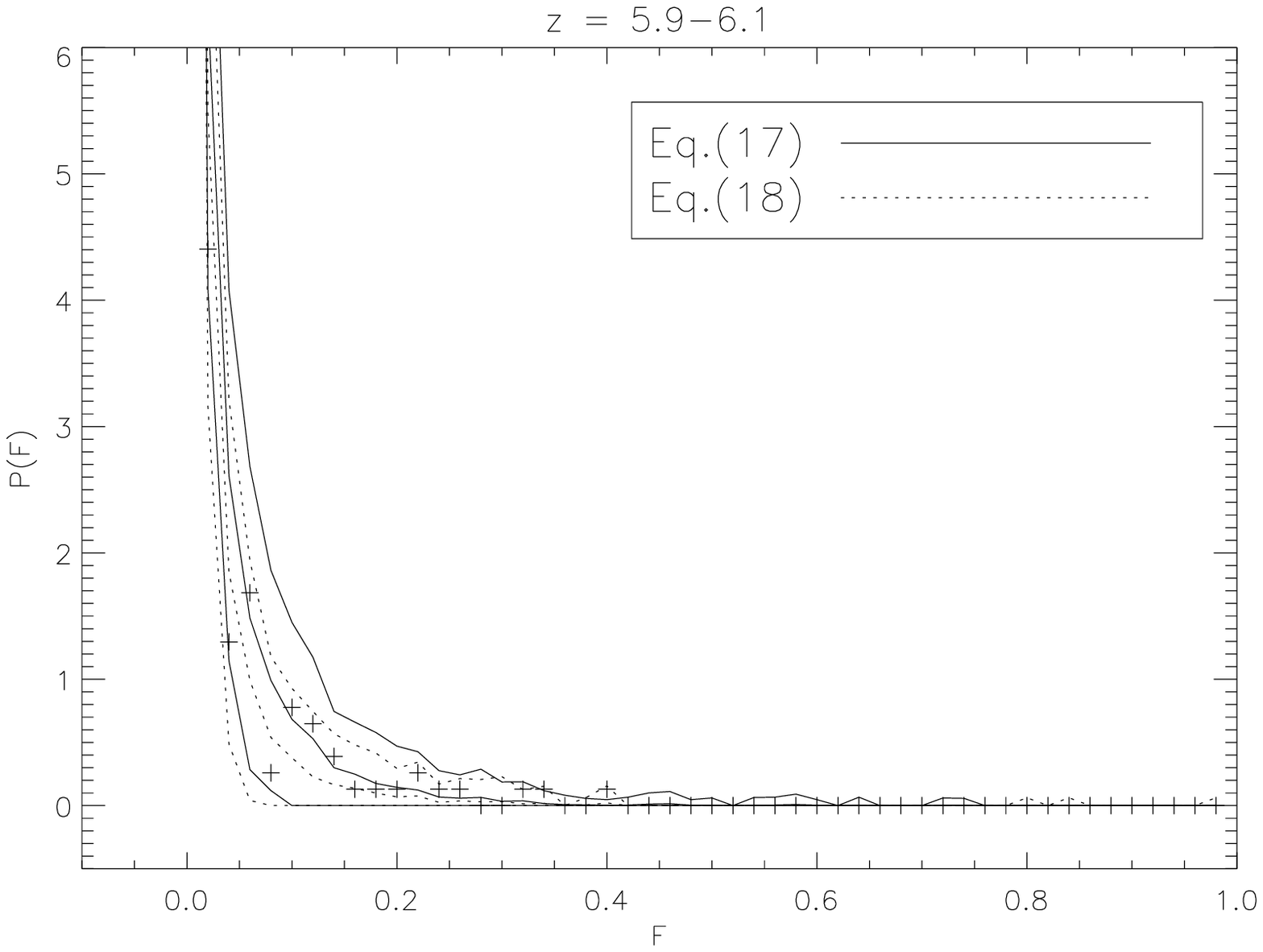}
\end{center}
\caption{Probability distribution functions of the transmitted flux at
$z=5.3-5.5$ (top left), $z=5.5-5.7$ (top right),$z=5.7-5.9$ (bottom
left), $z=5.9-6.1$ (bottom right). Cross are observation data. The solid
and dashed line in the center are results of samples generated with eq.(17)
and eq.(18) respectively. The above and below set of lines gives
the 1 $\sigma$ errors. }
\end{figure*}

Assuming ionization equilibrium under the uniform UV background, we
generate the field of neutral hydrogen fraction $f_{\rm HI}({\bf
x})=n_{\rm HI}(\bf x)/n_{\rm H}(\bf x)$ at each cell, where $n_{\rm
H}(\bf x)$ and $n_{\rm HI}(\bf x)$  are, respectively, the number
densities of hydrogen and neutral hydrogen  at ${\bf x}$. We
synthesize 160 samples of normalized Ly$\alpha$ transmitted flux
$\mathbb{F}(z)=\exp[-\tau(z)]$ from $z=4.9$ to $z=6.3$ using the
same methods as Zhang et al (1997) and Paschos \& Norman (2005). The
optical depth $\tau(z)$ is given by
\begin{equation}
\tau(z)=\frac{\sigma_0 c}{H}\int_{-\infty}^{\infty} n_{\rm HI}(x)
V[z-x-v(x), b(x)]dx,
\end{equation}
in which $\sigma_0$ is the effective cross-section of the resonant
absorption and $H$ is the Hubble constant at corresponding redshifts
of the samples. The Voigt function is $V[z-x-v(x),
b(x)]=1/(\pi^{1/2}b)\exp\{-[z-x-v(x)]^2/b^2(x)\}$, where $b(x)$
gives the thermal broadening.

Along an randomly selected lines of sight, we synthesize an
absorption spectrum  $\mathbb{F}$ from $z=4.9$ to $6.3$ by dividing
the spectrum into redshift intervals of $\Delta z=0.1$. As the
corresponding physical length scale for this redshift interval is
larger than our simulation box size, we integrate eq. (20) over the
simulation dump periodically. Each spectrum is resolved with the
same resolution of observation. Gaussian noise is added with
signal-to-noise ratio $S/N =10$.

We calculate the Gunn-Peterson optical depth and its dispersion of
the $160$ synthesized samples. The results are shown in Figure 8.
The observed results (Fan et al. 2006) are also shown in the Figure.
The simulation samples basically are consistent with the
observations except the dispersion at redshift $z=6.2$.  The
deviation at redshift $z=6.2$ might be due to the available observed
samples at the redshift range $z\sim6.2$ is too few.

We also do the statistics of the probability distribution function
of the transmitted flux  $\mathbb{F}(z)$ of observation and
simulation from $z=5.3$ to $z=6.1$. The results are presented in
Figure 9. The PDF of simulation samples can also fit observations
within 1-sigma range. Thus, all low order statistics of the
simulated samples are consistent with observation.

\subsection{Redshift dependence of $\beta$}

\begin{figure*}
\center
\includegraphics[scale=0.40,angle=-90]{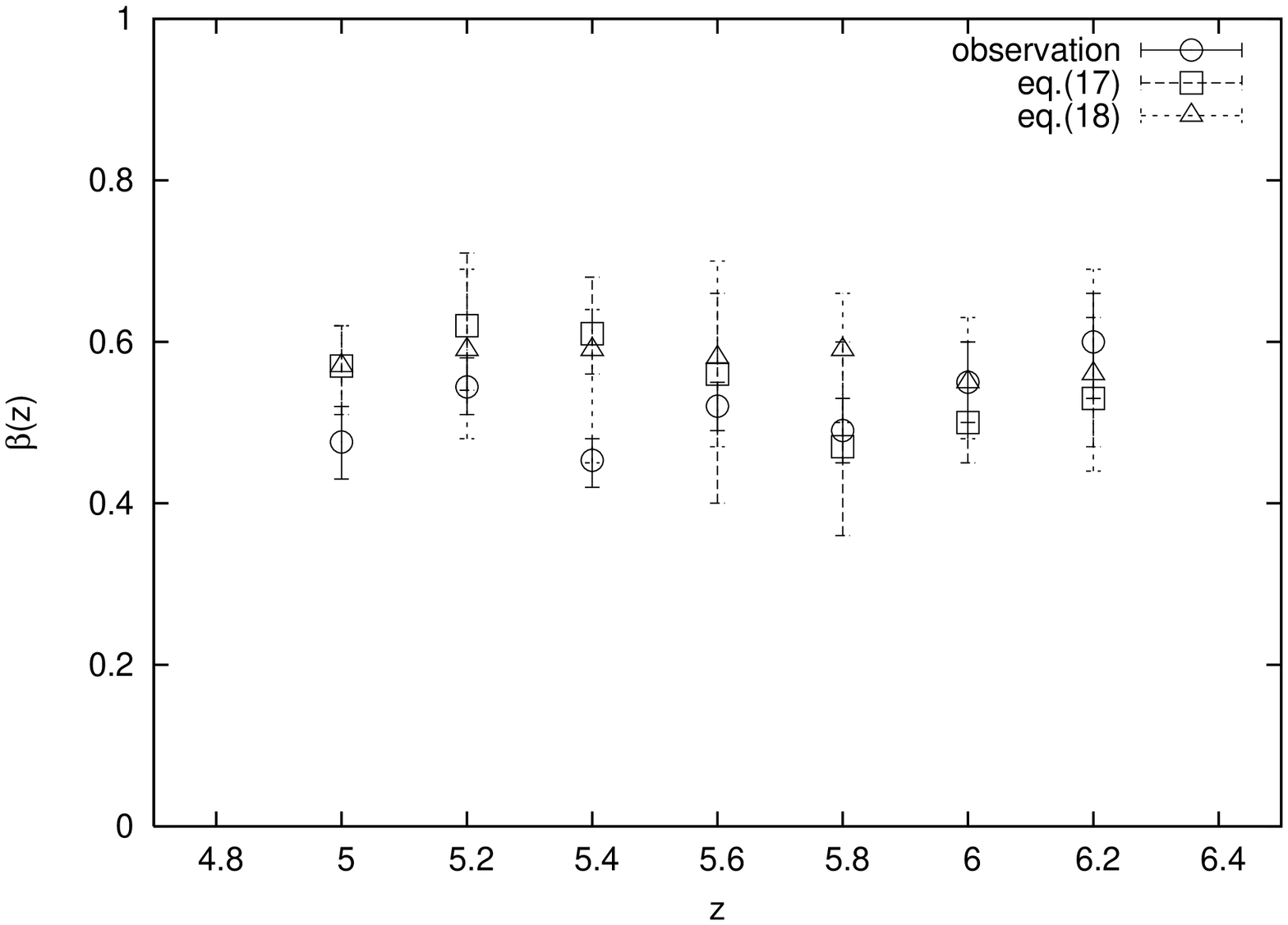}
\caption{Parameter $\beta$ vs. redshift for real data of Ly$\alpha$
transmitted flux (\S 3.2)(circle); and hydrodynamic simulation results
samples with  UV background eq.(17) (square), and eq.(18) (trigon).}
\end{figure*}

Using the method of \S 3.2, we calculate the parameters $\beta$ of simulated
samples of Ly$\alpha$ spectra.  To mimic the observation samples, which contains
totally 19 lines, we divide 160 one-dimensional simulation samples into 8
subsamples, each of which has 20 1-D samples. The uniform UV background eq.(17) and eq.(18)
are used to produce the simulation samples. Figure 10 presents the
redshift dependence of parameter $\beta$ for observed data and simulation samples.
The error bars of the simulation results are given by the maximum and minimum of
the 8 subsamples.

Although the redshift-dependence of the UV background eqs.(17) and
(18) are different at $z>5$, Figure 10 shows that the
redshift-dependence of $\beta$ given by the two models are about the
same. It is because the uniform field $J_{21}$ would not change the
non-Gaussianity significantly. The simulation results of parameter
$\beta$ are basically consistent with the observation, only at
$z=5.4$ shows a small deviation. Therefore, the $\Lambda$CDM
universe embedded with a uniform UV background history is able to
explain the feature of $\beta$ redshift-revolution.

Figure 7 shows that the parameters $\beta$ of neutral hydrogen
density field is in the range $0.30 - 0.35$, which is much less than
the results shown in Figure 10. It indicates that the
non-Gaussianity of neutral hydrogen field is different from that of
Ly$\alpha$ transmitted flux. This is because the non-Gaussianity of
Ly$\alpha$ transmitted flux depends not only on  the mass density
field of neutral hydrogen, but also on the velocity field via the
Voigt convolution eq.(20).

\subsection{Parameter $\gamma$}

Similar to \S 3.3, we use high order moment $\langle
\delta\tau_r^{2p}\rangle$ to quantify $K_p$ (eq.(15)) and then the
parameter $\gamma$. The results are listed in Table 2. The values of
$\gamma$ for $p=2$ and $3$ in each $z$ range are the same within
their errors. Comparing Table 1 and 2, the parameter $\gamma$ given
by observation data and simulation sample are also consistent with
each other within their errors.

\begin{table}
\center{Table 2. Parameter $\gamma$ at redshift $z=5.0$ to 6.2}
\vspace{-5mm}
\begin{center}
\bigskip
\begin{tabular}{l|ccc}
    \hline
    z           & $p$    & $K_p$   & $\gamma$     \\
        \hline
    5.0 & 2  & -0.29$\pm$ 0.04 & $0.27^{+0.04}_{-0.04}$\\
       &  3 & -1.71$\pm$ 0.06 &  $0.29^{+0.02}_{-0.03}$   \\
    5.4 & 2  & -0.29$\pm$ 0.06 & $0.29^{+0.06}_{-0.05}$\\
       &  3 & -0.73$\pm$ 0.12 &  $0.30^{+0.05}_{-0.04}$   \\
    5.8 & 2  & -0.23$\pm$ 0.03 & $0.20^{+0.03}_{-0.03}$\\
       &  3 & -0.59$\pm$ 0.07 &  $0.23^{+0.03}_{-0.03}$   \\
    6.2 & 2  & -0.15$\pm$ 0.11 & $0.14^{+0.10}_{-0.10}$\\
       &  3 & -0.38$\pm$ 0.24 &  $0.15^{+0.09}_{-0.10}$   \\
     \hline
\end{tabular}
\end{center}
\end{table}

The parameter $\gamma$ for observational samples of Ly$\alpha$
transmitted flux at $z=2.5$ is found to be $0.58\pm 0.20$ (Lu et al
2009), which is higher than that listed in Table 2. It might
indicate that parameter $\gamma$ is decreasing with redshift.
However, we should keep in mind that the error bars of observational
sample are large both at high and low redshift, the redshift
evolution of parameter $\gamma$ is not certain yet.

Figure 6 reveals that parameter $\gamma$ of neutral hydrogen is in
the range $0.80 - 0.89$, which is much larger than that shown in
Tables 1 and 2. Parameter $\gamma$  measures the singular structures
and higher $\gamma$ represents stronger singularity (Liu \& Fang,
2008; Lu et al 2009). Therefore, the field of neutral hydrogen mass
density contains much more singular structures than that of
Ly$\alpha$ transmitted flux.

\subsection{Spatially non-uniform UV background}

Whether the UV background at redshift $z \sim 5 - 6$ is spatially
uniform is an important problem of the history of reionization. With
only the redshift evolution of the optical depth and its variance at
$z\sim 5 - 6$, it seems to be difficult to distinguish models with
uniform background from inhomogeneous one. We will explore whether the 
log-Poisson non-Gaussian feature is dependent on the inhomogeneity of 
the UV background.

\begin{table*}
\center{Table 3. Parameter $\beta$ and UV background models}
\vspace{-5mm}
\begin{center}
\bigskip
\begin{tabular}{l|ccccc}
    \hline
    z       & A    & B   & C & D    \\
        \hline
5.4 & 0.48$\pm$ 0.02  & 0.57$\pm$ 0.02 & 0.56$\pm$ 0.02 & 0.55$\pm$ 0.02 \\
5.8 & 0.52$\pm$ 0.02  & 0.50$\pm$ 0.02 & 0.53$\pm$ 0.02 & 0.52$\pm$ 0.02 \\
     \hline
\end{tabular}
\end{center}
\end{table*}

Many works have been done on the fluctuation of hydrogen-ionizing
radiation background field. Analytical method and Monte Carlo
simulations with randomly distributed sources were firstly used to
estimate the intensity fluctuations and its effect on the Ly$\alpha$
forest (Zuo 1992; Fardal \& Shull 1993; Croft et al. 1999). Large
scale N-body simulations and hydrodynamical simulations then were
applied to study the fluctuations of the UV background (Gnedin \&
Hamilton 2002; Meiksin \& White 2003; Meiksin \& White 2004;Croft
2004; Bolton et al. 2006). Investigations have been carried out recently 
on this problem at high redshift (Wyithe \& Loeb 2005; Mesinger \& Furlanetto, 2009;
Furlanetto \& Mesinger 2009). The general features of an inhomogeneous UV background
can be sketched as
\begin{equation}
J_{21}({\bf x}, z)=\bar{J}_{21}(z)[1+\delta({\bf x},z)],
\end{equation}
where $\bar{J}_{21}(z)$ is the UV background given by eqs.(17) or
(18). Equation (21) means, the field of the UV background is
fluctuated with respect to its mean $\bar{J}_{21}(z)$.

Unfortunately, no commonly accepted $J_{21}({\bf x}, z)$ or
$\delta({\bf x},z)$ are available. The fluctuation field
$\delta({\bf x}, z)$ may or may not be correlated with the density
distribution of cosmic baryon matter. In this context, we consider 4
toy models of $\delta({\bf x})$ as follows:
\begin{itemize}
\item A. $\delta({\bf x})=0.1g({\bf x})$, where $g({\bf x})$ is a Gaussian random
field with variance $\sigma^{2}=1.0$.
\item B. $\delta ({\bf x})=0.3g({\bf x})$.
\item C. $\delta({\bf x})$ is given by
\begin{equation}
\delta({\bf x})=\left\{  \begin{array}{ll}
    \delta_0 \sin(\frac{\pi}{2} \min(\lg(\rho({\bf x}))/2.0,1.0))   &  \rho({\bf x})>1.0 \\
    \delta_0 (\frac{\pi}{2} \max(\lg(\rho({\bf x}))/2.0,-1.0)) &  \rho({\bf x})<1.0
\end{array}  \right.
\end{equation}
where $\delta_0=0.1$.
\item D. The same as model C, but $\delta_0=0.3$.
\end{itemize}
For models A and B, the fluctuations of UV background are
statistically independent of the IGM density field, while it is
correlated to the IGM distribution for models C and D.

Because the radiative transfer has not yet been included in our
simulation, we simply add the non-uniform UV background to the
outputs of simulations with eq.(17) and produce samples of the
neutral hydrogen density field and Ly$\alpha$ transmitted flux at
redshift 5.4 and 5.8. Since our goal is only to see whether the
non-Gaussianity is affected by a fluctuating UV background, the post
processing method would be acceptable as a first try.

The $\beta$ values of the Ly$\alpha$ transmitted flux samples based
on non-uniform UV background eq.(22) are listed in Table 3. Although the 
$\beta$ values among models A,B,C and D sometimes show 1-$\sigma$ deviation 
from each other, these simplified models seem to have no effect in general. 
It may indicate that the effect of inhomogeneous UV background on the 
log-Poisson non-Gaussian feature is not detectable yet with current 
data.

\section[]{Discussion and conclusion}

Nonlinear evolution of baryon fluid at high redshift is a central
problem of cosmology. It is well known that in the nonlinear regime
the dynamical behavior of cosmic baryon fluid does not always trace
the collisionless dark matter. The non-Gaussianity of the mass and
velocity fields of baryon fluid is given by the hydrodynamics of the
cosmic flow. A common property of the evolution of a Navier-Stokes
fluid is to reach a turbulent state when the Reynolds number is high
(e.g. Zhu et al. 2010). In the scale free range, the fully developed
turbulence is of statistically quasi-steady characterized by
log-Poisson hierarchy. In this regime, the cosmic baryon fluid
undergoes the evolution of clustering and finally falls into massive
halos of dark matter to form structures, including light-emitting
objects.

The cosmic baryon fluid or the IGM, both at high and low redshifts,
exhibits the log-Poisson non-Gaussian features in the range from the
onset scale of the nonlinear evolution to the dissipation scale,
i.e. the Jeans length. The log-Poisson non-Gaussianity has been
identified in various simulated and observed samples of the IGM,
including the mass density and velocity fields of baryonic matter, the
density field of neutral hydrogen, and Ly$\alpha$ transmitted flux.
Although these fields are different from each other, they show the
common behavior of log-Poisson non-Gaussianity. This scenario is
further supported with high redshift Ly$\alpha$ transmitted flux.

The log-Poisson non-Gaussian parameter $\beta$ of the mass density
field on the same physical scales is found to be increasing with
redshift (Liu \& Fang 2008). It implies that the mass density field
of the IGM is less non-Gaussian at high redshifts. The log-Poisson
non-Gaussianity at high redshifts should be weak than that of low
redshifts on the same physical scale. However, we found in this
paper that observed data of Ly$\alpha$ transmitted flux at redshift
$z= 5$ - 6 shows about the same level of log-Poisson non-Gaussianity 
as low redshift $z\sim 2$ data. This is because the scales covered by
the data at $z=5$ - 6 are smaller than the data at $z\sim 2$. The
weak evolution of $\beta$ given by this paper implies that
turbulence state of cosmic baryon fluid at $z\sim 2$ on physical
scales 1 - 10 h$^{-1}$ Mpc is about the same as that at $z= 5$ - 6
on physical scales 0.1 - 1 h$^{-1}$ Mpc. This is reasonable 
considering that the Jeans length and the typical scale of onset of
nonlinear evolution are increasing with time. The log-Poisson
non-Gaussianity at low and high redshifts are about the same
once the nonlinear evolution is fully developed.

In a word, on the scales of fully developed turbulence, the
parameter $\beta$ of the IGM at high and low redshifts should be
about the same. This property is very different from the
Gunn-Peterson optical depth and its variance, both show a
strong redshift dependence. The Gunn-Peterson optical depth and its
variance are sensitive to the intensity of the UV background. 
Oppositely, the parameters of the log-Poisson non-Gaussianity are
weakly dependent on the intensity of the UV background. A more self-consistent 
and delicate handle of the inhomogeneity of 
the UV background in the cosmic hydrodynamic simulation and much more
high quality observation data would help us to use the log-Poisson 
non-Gaussianity to investigate the ionizing background at high redshift.

\section[]{Acknowledgments}

WSZ acknowledges the support of the International Center for Relativistic Center
Network (ICRAnet).

\end{document}